\documentclass[twocolumn]{aastex63}
\usepackage{graphicx}
\usepackage{xcolor}

\newcommand{\ztf}{ZTF18aazmehw}

\newcommand{\swift}{Swift J0503.7-2819}
\newcommand{\igr}{IGR~J19552+0044}
\newcommand{\rxs}{1RXS~J083842.1$-$282723}

\newcommand{\ratio}{$P_{spin}/P_{orb}$}
\newcommand{\cpd}{cycles~d$^{-1}$}

\newcommand{\gaia}{Gaia21akb}

\begin{document}

\title{TESS light curves of two new magnetic cataclysmic variables: an asynchronous polar at the period minimum, and an eclipsing system with a large spin-to-orbit ratio}

\author[0000-0001-7746-5795]{Colin Littlefield}
\affiliation{Bay Area Environmental Research Institute, Moffett Field, CA 94035, USA}

\author[0000-0002-4005-5095]{Krystian I\l{}kiewicz}
\affiliation{Nicolaus Copernicus Astronomical Center, Polish Academy of Sciences, Bartycka 18, 00-716 Warsaw, Poland}

\author[0000-0002-5897-3038]{Paul A. Mason}
\affiliation{New Mexico State University, MSC 3DA, Las Cruces, NM, 88003, USA}
\affiliation{Picture Rocks Observatory, 1025 S. Solano Dr. Suite D., Las Cruces, NM 88001, USA}

\author[0000-0003-4069-2817]{Peter Garnavich}
\affiliation{Department of Physics, University of Notre Dame, Notre Dame, IN 46556, USA}

\author[0000-0001-5387-7189]{Simone Scaringi}
\affiliation{Centre for Extragalactic Astronomy, Department of Physics, Durham University, South Road, Durham, DH1 3LE}
\affiliation{INAF -- Osservatorio Astronomico di Capodimonte, Salita Moiariello 16, I-80131 Naples, Italy}

\correspondingauthor{Colin Littlefield}
\email{clittlef@alumni.nd.edu}

\begin{abstract}

   A recent development in the study of magnetic cataclysmic variable stars (mCVs) has been the identification of asynchronously spinning mCVs with orbital periods $<2$~h that have significantly higher white dwarf spin-to-orbital period ratios than their longer-period counterparts. We report the discovery of two additional mCVs in this class. The first, Gaia21akb, is a candidate asynchronous polar at the period minimum. While TESS photometry cannot, in isolation, lead to a conclusive identification of the orbital period, the probable orbital period of 1.29~h would be the second-shortest of any known polar and would result in a spin-to-orbit ratio of 0.9879. The second system in our study, ZTF18aazmehw, is an eclipsing mCV with a 1.50~h orbital period and a spin-to-orbit ratio of 0.867. Contrary to expectations for an asynchronous polar, ZTF18aazmehw does not show discernible evidence of pole switching and might possess a disk-like structure. The increasing number of short-period asynchronous mCVs with large spin-to-orbit ratios lends credence to theoretical predictions that asynchronously rotating mCVs with sufficiently strong white dwarf magnetic fields can achieve synchronization when their orbital separations have shrunk sufficiently.

\end{abstract}

\section{Introduction}

    Magnetic cataclysmic variable stars (mCVs) are short-period binaries containing an accreting, magnetized white dwarf primary. If the field strength is high enough ($\gtrsim3$~MG), the WD's magnetic field synchronizes the WD's rotation to the binary orbit. Synchronously rotating systems are known equivalently as polars or as AM~Herculis stars. See the comprehensive book by \citet{Warner1995cvs..book.....W} for a review of both mCVs and non-magnetic CVs.

    Asynchronous rotation is commonly observed in several classes of mCVs, of which the asynchronous polars (APs) are the closest to being synchronous. As their name implies, APs share many characteristics with polars except that their spin and orbital periods differ by up to several percent. The prototype AP, V1500~Cyg, was identified as it declined from its 1975 nova eruption \citep{stockman88}. V1500~Cyg is widely believed to have been a synchronous polar before its nova, and measurements of its spin-period derivative show that the WD is approaching synchronization \citep{schmidt91, pavlenko_v1500cyg}. However, there is currently scant evidence of a recent nova eruption in the other APs, and searches for nova shells around other APs have yielded null results \citep{pagnotta16, sd22}.

    At the other extreme, intermediate polars (IPs) are  highly asynchronous \citep{patterson94}, with common spin-to-orbit ratios of \ratio\ $\lesssim 0.1$, especially above the period gap. IPs are often assumed to have lower magnetic field strengths than polars, and because of that, their asynchronous rotation is an equilibrium state.

    Until recently, there was a relatively clear-cut distinction between polars, asynchronous polars, and IPs. IPs generally had spin-to-orbit ratios far below unity, while APs had \ratio\ $\gtrsim 0.95$. Only two IPs, EX~Hya
    \citep[\ratio$=0.68$; ][]{ex_hya} and Paloma
    \citep[\ratio$=0.87$; ][]{schwarz_paloma, littlefield_J0846}, deviated from this tendency.  However, a significant development in recent mCV research has been the identification of a population of systems with a theoretically challenging combination of characteristics: (1) orbital periods under the 2~h period gap and (2) $0.5\lesssim$\ratio$\lesssim0.9$, placing them in a previously sparse parameter space. These systems could be interpreted, somewhat confusingly, as very de-synchronized APs (if the asynchronous rotation is a temporary disequilibrium in a previously synchronized system) or as nearly synchronous IPs (if it is an equilibrium condition). To maintain consistency with earlier literature, we use the term ``AP'' to refer to asynchronously rotating mCVs that are within several percent (\ratio $\gtrsim 0.95$) of being synchronous. However, we note that the terms ``IP'' and ``AP'' emerged in an era when it appeared that very few mCVs had $0.5 \lesssim $\ratio$ \lesssim 0.95$; in those simpler times, the measurement of \ratio\ was therefore sufficient to classify an mCV within this dichotomy. Now that it is becoming clear that mCVs with $P_{orb} < 2$~h can have any \ratio, we suggest that the term ``asynchronous mCV'' might be preferable because it avoids any implicit assumptions about the rotational equilibrium of the WD as well as its pre-discovery behavior.

    One reason for the recent spate of newly discovered asynchronous mCVs in the compilation in Table~\ref{table:APs} is the availability of nearly uninterrupted, high-cadence light curves from space-based instruments, supplemented by ground-based survey photometry with deep limiting magnitudes. Indeed, many of the newly identified systems in Table~\ref{table:APs} have been hiding in plain sight, having been misclassified for years as garden-variety systems. Perhaps the most extraordinary example of this is V844~Her, which was once described as a textbook example of an SU~UMa-type dwarf nova \citep{Kato_V844her}. Nevertheless, \citet{v844her_paper1} detected a 29-min period during a superoutburst in its Transiting Exoplanet Survey Satellite \citep[TESS; ][]{tess} light curve suggestive of an asynchronously rotating magnetic WD; subsequent time-series spectroscopy revealed that this periodicity is also present in the system's high-velocity Balmer and He~I emission, leading Greiveldinger et al. (AJ, submitted) to classify it as an IP with \ratio\ of either 0.373 or 0.746 (most likely the former). Likewise, three of the systems in Table~\ref{table:APs}, FR~Lyn \citep{fr_lyn}, SDSS J084617.11+245344.1 \citep{littlefield_J0846}, and SDSS J134441.83+204408.3 \citep{littlefield_J1344} had originally been misclassified as synchronous polars; however, data from the Zwicky Transient Facility \citep[ZTF;][]{ztf_paper}, Kepler K2 \citep{howell_k2}, and TESS, respectively, established that each of these systems is actually asynchronous. In a similar vein, another system in Table~\ref{table:APs}, Swift~J0503.7-2819, had been misclassified as a fairly standard IP until a pair of studies in 2022 led to the realization that \ratio $=0.79$ \citep{halpern22, rawat22}. Against this backdrop, it seems probable that there are additional mCVs with \ratio$>0.6$ that have been similarly misclassified.

   Here, we analyze the TESS photometry of two understudied mCVs, \gaia\ and \ztf, whose coordinates are provided in Table~\ref{table:astrometry}. We show that both of these obscure systems are actually asynchronous mCVs with remarkably large \ratio\ ratios.

    \begin{deluxetable*}{ccccccc}
            \tablecaption{The asynchronous mCVs with \ratio$>$~0.6 \label{table:APs}}

            \tablehead{
                \colhead{Name} &
                \colhead{$P_{orb}$ (h)} &
                \colhead{$P_{spin}/P_{orb}$} &
                \colhead{$P_{beat}$ (d)} &
                \colhead{Distance (pc)} &
                \colhead{References}
                }

            \startdata
            \gaia     &  1.29  &  0.9879   &   4.41  & $358^{+13}_{-12}$     & this work          \\
            \swift    & 1.36 & 0.79  & 0.217& $837^{+60}_{-43}$     & \citet{halpern22, rawat22}\\
            \igr      & 1.39 & 0.972 & 2.04 & $165.5^{+1.9}_{-1.5}$ & \citet{tovmassian} \\
            \ztf      &  1.50  &   0.867   &   0.405   &    $504^{+24}_{-27}$        & this work                  \\
            ZTF18abaaewz & 1.55 & 0.61 & 0.100 & $510^{+140}_{-110}$ & \citet{szkody24}\\
            EX Hya & 1.638 & 0.68 & 0.146 & $56.77\pm0.05$ & \citet{ex_hya}  \\
            \rxs      & 1.640 & 0.952 & 1.34 & $156.0^{+1.9}_{-2.2}$ & \citet{halpern24} \\
            CD Ind    & 1.87 & 0.989 & 7.3  & $235.3^{+4.0}_{-3.2} $ & \citet{littlefield, mason20}\\
            FR Lyn & 1.89 & 0.9968 & 24.6 & $478^{+48}_{-42}$ & \citet{fr_lyn}\\
            SDSS J134441.83+204408.3 & 1.90 & 0.893 & 0.658 & $599^{+53}_{-46}$ & \citet{littlefield_J1344}\\
            Paloma    & 2.62 & 0.87 & 0.71 & $582^{+28}_{-20}$ & \citet{schwarz_paloma, littlefield_J0846} \\
            V1500 Cyg & 3.351 & 0.986 & 9.58 & $1570^{+270}_{-190} $ &\citet{pavlenko_v1500cyg}\\
            BY Cam    & 3.354 & 0.99  & 15  & $264.5^{+1.9}_{-1.7}$ & \citet{mason22}\\
            V1432 Aql & 3.366 & 1.002 & 62 & $450\pm7$ & \citet{littlefield15}\\
            SDSS~J084617.11+245344.1 & 4.64 & 0.972 & 6.77 & $1230^{+800}_{-290}$ & \citet{littlefield_J0846}\\
            \enddata

            \tablecomments{An earlier version of this table appeared in \citet{littlefield_J0846}, but this version has been updated with several new systems. \citet{hakala_cd_ind} and \citet{wang2020} propose two alternate sets of frequency identifications for CD~Ind; \citet{wang2020} also contends that BY~Cam's frequencies have been misidentified. Another system, V844~Her, has a likely \ratio\ of 0.373 \citep{v844her_paper1}. ZTF18abaaewz is referred to as 1631+69 in shorthand in \citet{szkody24}, while \citet{fr_lyn} use FR~Lyn's SDSS identifier. We provide the geometric distances computed by \citet{BJ21} from Gaia eDR3 \citep{edr3}.}

        \end{deluxetable*}

    \begin{deluxetable*}{ccccccc}
        \label{table:astrometry}
        \tablecaption{Gaia DR3 astrometry for \gaia\ and \ztf}
        \tablehead{
            \colhead{Name} &
            \colhead{Gaia DR3 identifier} &
            \colhead{Epoch} &
            \colhead{$\alpha$} &
            \colhead{$\delta$} &
            \colhead{$\mu_{\alpha}$} &
            \colhead{$\mu_{\delta}$} \\
            \colhead{} & \colhead{} & \colhead{}
            & \colhead{} & \colhead{} & \colhead{mas yr$^{-1}$} & \colhead{mas yr$^{-1}$}
        }
        \startdata
        \gaia  & 2395305769240905600 & 2016 & 23h29m26.09039s & -16d16m34.5686s & -14.8 & -14.6\\
        \ztf & 4321588332240659584 & 2016 & 19h25m30.53281s & +15d54m26.3697s & -16.9 & -12.3 	\\
        \enddata
    \end{deluxetable*}

\section{\gaia: a candidate asynchronous polar at the period minimum}

    \begin{figure*}
        \includegraphics[width=\textwidth]{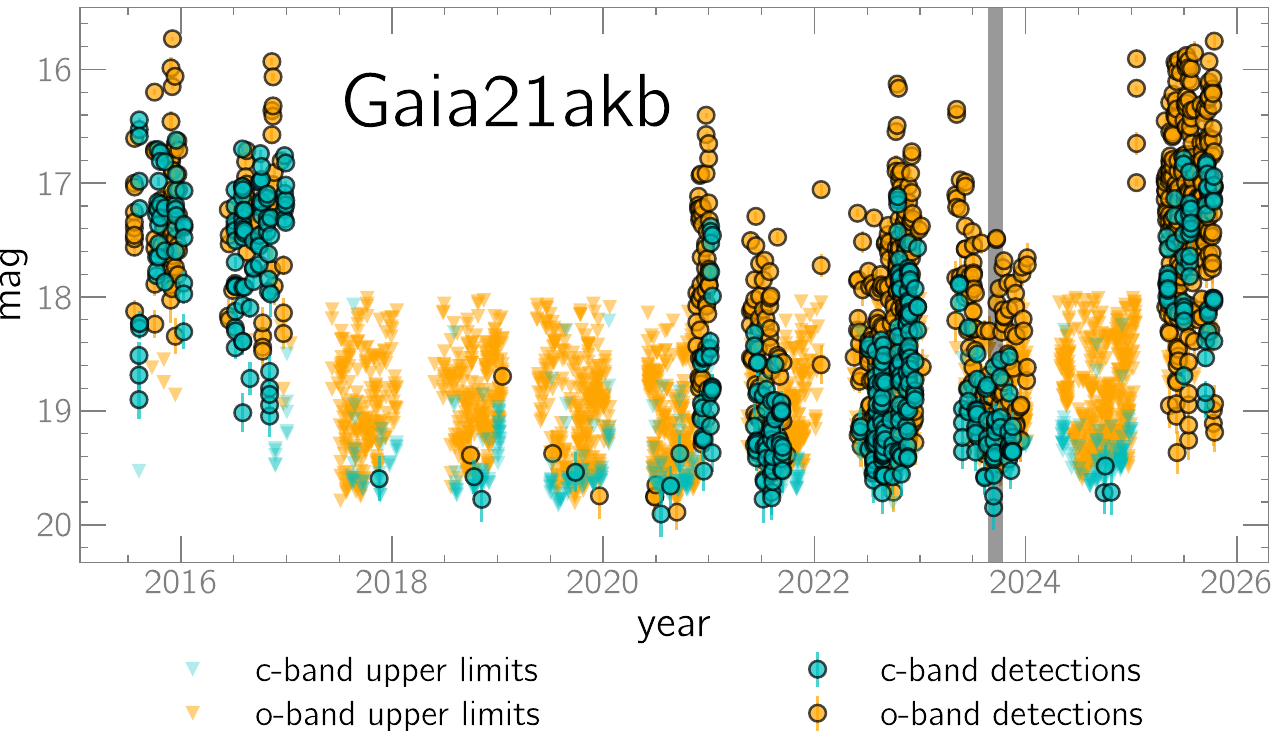}
        \caption{ATLAS light curve of Gaia21akb, with the time of the TESS observations indicated in gray. For clarity, upper limits brighter than magnitude 18.0 are not shown. The data are consistent with the behavior of polars, which often alternate unpredictably between bright states and low-accretion states. \label{fig:atlas_gaia}
        }
    \end{figure*}

    \subsection{Overview}

    The previous literature on \gaia\ can be summarized succinctly: there is none. The AAVSO International Variable Star Index catalog \citep[VSX; ][]{watson} classifies it as an AM Herculis (polar) system based on Taichi Kato's discussion of its ZTF light curve in vsnet-chat~8727,\footnote{http://ooruri.kusastro.kyoto-u.ac.jp/mailarchive/vsnet-chat/8727} but there is no published work examining this system. The distance to \gaia\ is $358^{+13}_{-12}$~pc \citep{BJ21}, and the VSX lists some of its many aliases as ZTF20acqpkxj, AT~2020abey, and TIC~328010691.

    \subsection{ATLAS \& TESS photometry}

    Our dataset for \gaia\ consists of observations obtained with both the
    Asteroid Terrestrial-impact Last Alert System \citep[ATLAS; ][]{atlas} and TESS. We begin by analyzing the former.

    ATLAS photometry of \gaia\ began in 2015 and has continued until the present day (Fig.~\ref{fig:atlas_gaia}). These data show large-amplitude variability on multiple timescales. For example, in the years $2015-2016$ and $2022-2024$, the system was regularly bright enough to be detected, but between $2017-2020$, it experienced a prolonged low state. The high state at the beginning of the ATLAS data appears considerably brighter than subsequent high states, with a typical $c$-band magnitude of $\sim17$ compared to $\sim18-19$ in other high states. The observed alternation between high and low states is consistent with the behavior of many polars in long-term survey photometry \citep{mason15, duffy22}.

    The TESS observation of \gaia\ took place during sectors 69 and 70, near the end of the $2022-2024$ high state. The data were obtained at a cadence of 2 min, and because they overlap with ATLAS data, it is possible to use the ATLAS observations to flux-calibrate the TESS data. Although it is not strictly necessary to do so, flux-calibrating the TESS data can simplify the ensuing analysis. For example, a single TESS pixel subtends 21\arcsec, and the resulting dilution can make it difficult to reliably characterize the true amplitude of variation of the target.

    To accomplish the flux calibration of the TESS data, we applied a barycentric correction to the ATLAS timestamps and extracted all ATLAS $o$-band data that were obtained within $\pm$1~min of a TESS observation. We used only the ATLAS $o$-band data because it is a much better match to the red-infrared TESS bandpass than the much bluer $c$-band. For each sector, we calculated a linear regression to describe the TESS counts as a function of the simultaneous ATLAS flux density (in units of $\mu$Jy). The $y$-intercept of this function can be thought of as the predicted flux reported by TESS if the source were to fade to 0~$\mu$Jy. Meanwhile, the slope predicts how much the TESS count rate will increase should the source brighten by 1~$\mu$Jy in the $o$-band. We weighted the regressions by the signal-to-noise ratio of the ATLAS data and applied this procedure to each TESS sector individually because the pointing, the photometric apertures, and the blending vary across different TESS sectors.

    \begin{figure*}
        \includegraphics[width=\textwidth]{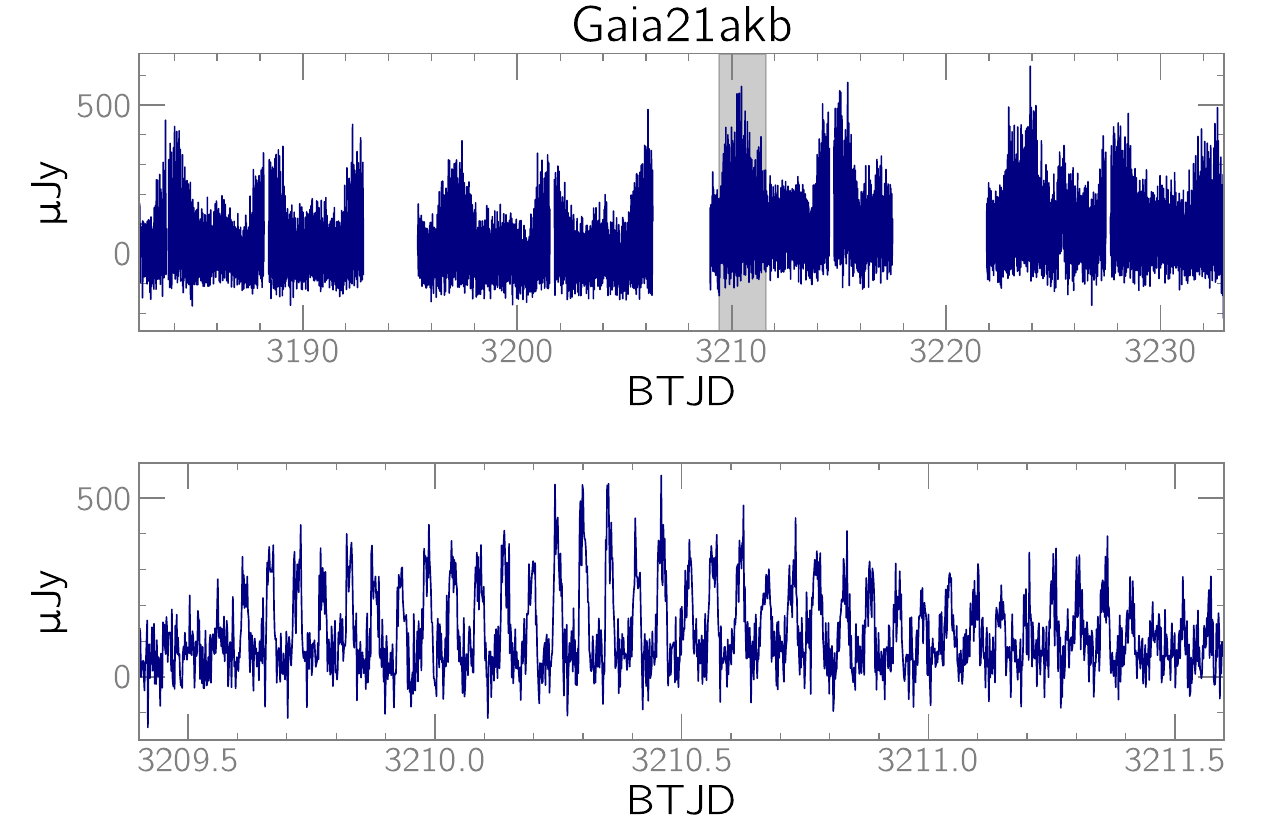}
        \caption{TESS light curve of \gaia\ from sectors 69-70. The data were flux-calibrated using simultaneous ATLAS $c$-band data as described in the text. The lower panel shows an enlargement of the short-period variability within the shaded region in the upper panel.
        \label{fig:gaia_tess_lc}
        }
    \end{figure*}

    \begin{figure*}
        \includegraphics[width=\textwidth]{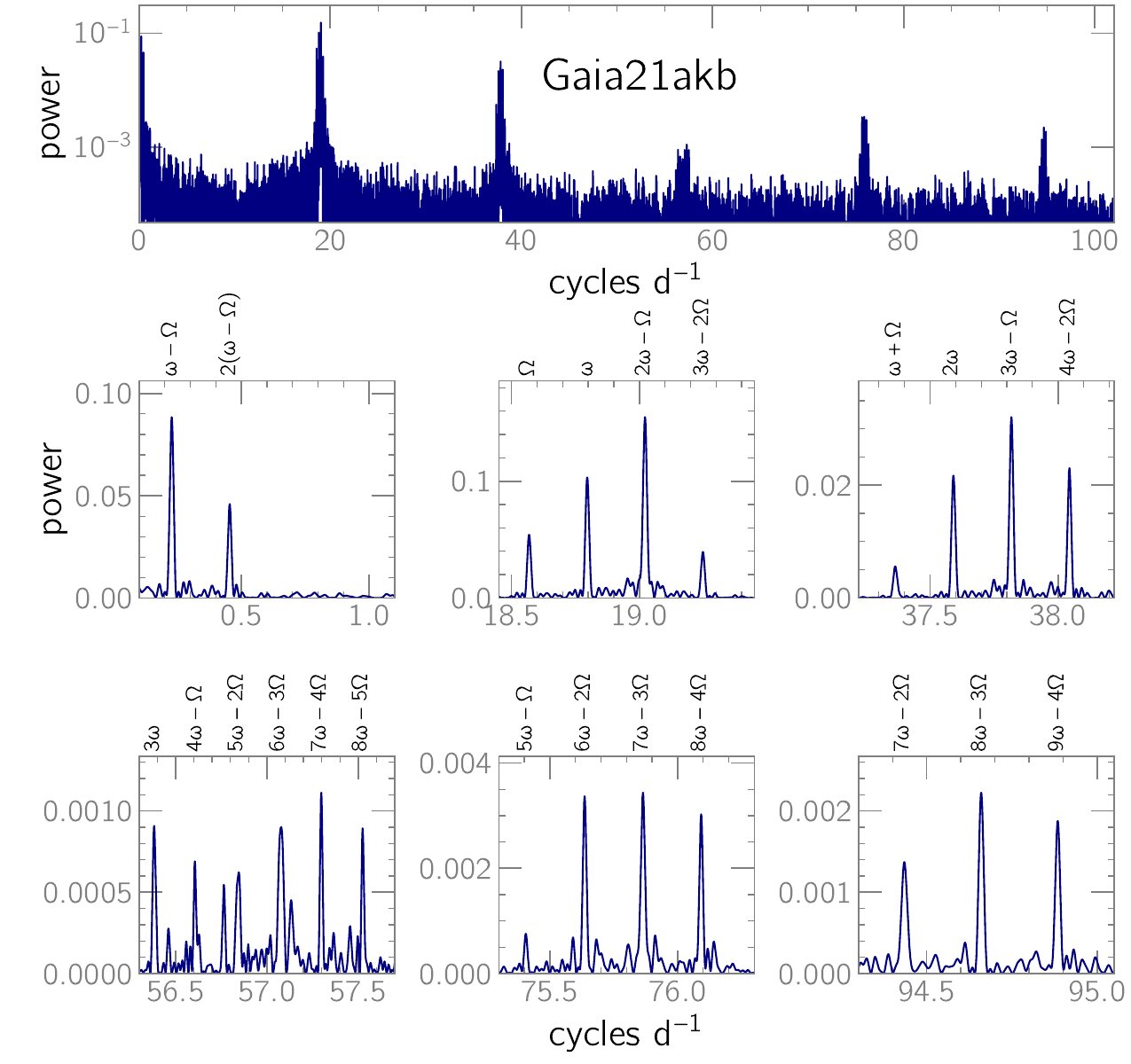}
        \caption{TESS power spectrum of \gaia. The top row shows the full power spectrum, while the six panels in the bottom two rows show enlarged segments of the power spectrum with our proposed frequency identifications. While there is little doubt as to the identity of the beat frequency $\omega-\Omega$ (middle row, left panel) and its second harmonic, the correct identifications of $\omega$ and $\Omega$ are comparatively uncertain for reasons discussed in the text.\label{fig:gaia21akb_power_1d}
        }

    \end{figure*}

    \begin{figure*}
        \includegraphics[width=\textwidth]{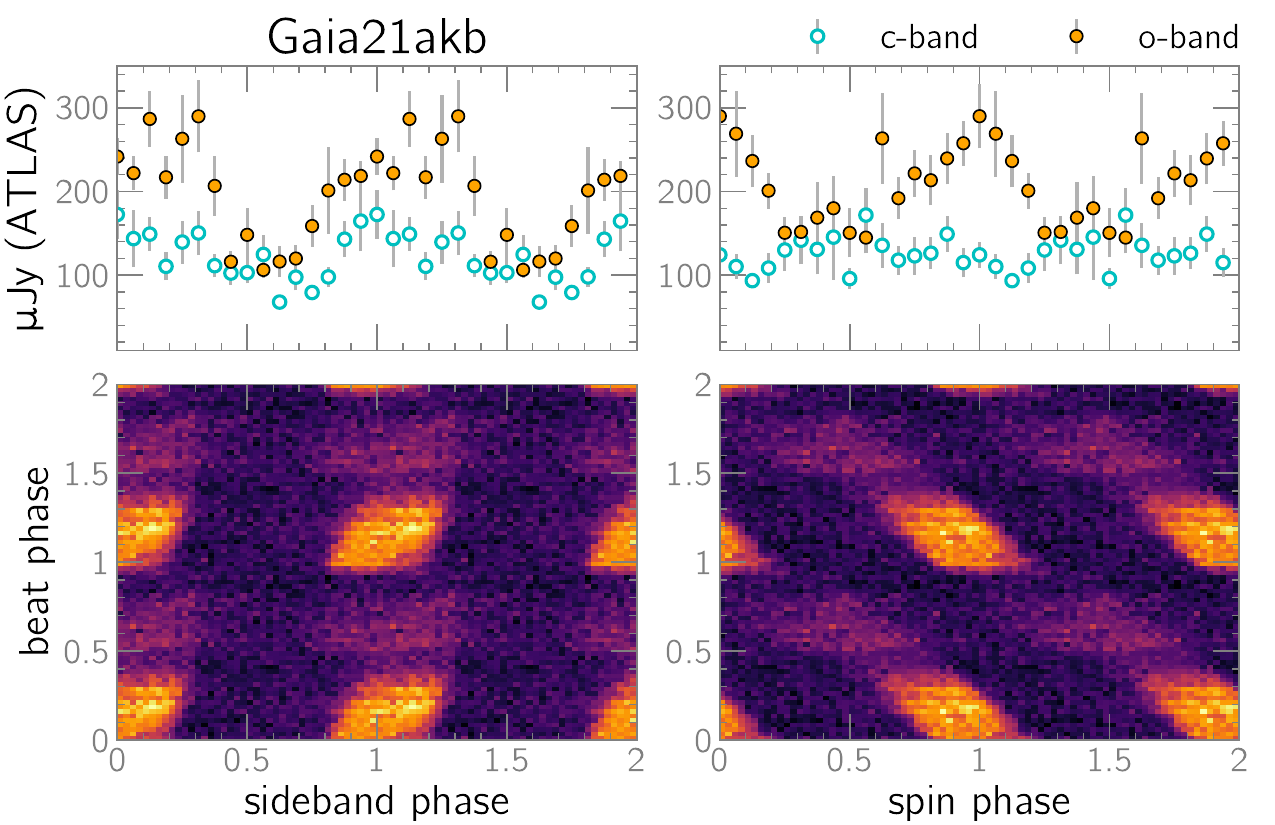}
        \caption{Light curves of \gaia, phased to our provisional frequency identifications. {\bf Top:} Phase-averaged 1D profiles as observed by ATLAS between 2022-2024. The amplitude of variation is significantly higher at longer wavelengths. {\bf Bottom:} 2D TESS light curves showing the evolution of the $2\omega-\Omega$ and $\omega$ profiles across the beat cycle. Our frequency identifications suggest that there are two accretion regions on opposite sides of the WD and that the accretion flow switches between them in opposite halves of the beat cycle. If the sideband frequency had been misidentified as the spin frequency, it would have appeared that both accretion regions were in the same longitudinal hemisphere of the WD.
        \label{fig:gaia_2d_lc}
        }
    \end{figure*}

    Fig.~\ref{fig:gaia_tess_lc} shows the resulting flux-calibrated light curve of \gaia. It is immediately obvious that the light curve contains a short-period variation whose amplitude undergoes cyclical changes on a $\sim4$-day timescale. A Lomb-Scargle power spectrum of that light curve (Fig.~\ref{fig:gaia21akb_power_1d}) reveals five distinct groups of high-frequency signals that are harmonically related. Within each of these clusters, there is a series of sideband frequencies  separated by 0.25~\cpd\ intervals. Moreover, that spacing between sidebands is itself directly visible in the power spectrum as a frequency at 0.25~\cpd\, along with its second harmonic. All told, there are over 20 frequencies that can be expressed as linear combinations of just two underlying frequencies.

    The power spectrum is very typical of an AP. While AP power spectra are notoriously complicated \citep{mason20}, they are expected to show a conspicuous low-frequency signal corresponding to the beat frequency ($\omega-\Omega$) between the WD spin ($\omega$) and binary orbital ($\Omega$) frequencies. In the case of \gaia, there are just two low-frequency signals, both of which are harmonically related with the fundamental at 0.25~\cpd, which we identify as $\omega-\Omega$.

    However, individually identifying both $\omega$ and $\Omega$ is significantly more challenging. There are four equally spaced signals between 18.5~\cpd and 19.3~\cpd, the strongest of which is at 19~\cpd, and any of these could be at least a plausible candidate to be the orbital frequency. Following the discussion in Sec.~3.6 of \citet{szkody24}, our interpretation of these signals is guided by three assumptions, each of which is debatable.

    \begin{itemize}
        \item   We presume that $\omega > \Omega$ because this is true for all known asynchronous mCVs except for the singular case of V1432~Aql.\footnote{As our referee pointed out, no AP except for V1432~Aql undergoes eclipses, so the long-assumed identity of $\Omega$ in those systems deserves careful consideration, and \citet{wang2020} proposed that $\Omega > \omega$ in CD~Ind and BY~Cam. However, in BY~Cam, $\Omega$ has been measured from the radial-velocity variations of emission from the secondary's inner hemisphere \citep{schwarz_bycam}, and in V1500~Cyg, the extreme reflection effect from the secondary's inner hemisphere enables the orbital period to be identified with photometry \citep{pavlenko_v1500cyg}.}

        \item When $\Omega$ is ambiguous, we favor identifications of $\Omega$ that result in periods at or above the period minimum for hydrogen-rich donors.

        \item We assume that the power spectrum will generally follow theoretical predictions.

    \end{itemize}

    Guided by these assumptions, we provisionally identify the 19.02~\cpd\ signal as $2\omega-\Omega$. This frequency, which results from pole switching, is expected to be the dominant signal in AP power spectra. This inference, in conjunction with our presumption that $\omega>\Omega$, leads to
    $\omega = 18.79569$~\cpd\ (76.61~min),
    $\Omega = 18.56873$~\cpd\ (77.55~min), and \ratio = 0.9879, consistent with an AP classification.

    A common method of analyzing the light curves of APs is to phase them to the beat frequency ($\omega - \Omega$), the frequency at which the magnetic field rotates with respect to any structures that are fixed in the binary rest frame, such as the ballistic accretion stream. Fig.~\ref{fig:gaia_2d_lc} shows how the $2\omega-\Omega$ and $\omega$ profiles evolve over the beat cycle. Notably, the spin pulse drifts backward in phase as the system progresses through the beat cycle, meaning that the interval between consecutive spin-pulse maxima is shorter than the actual spin period. \citet{gs97} showed that a phase drift of the spin profile is an expected consequence of asynchronous rotation in APs. As the WD rotates with respect to the accretion stream, the stream will be forced to couple to different field lines, causing the accretion footprint on the WD's surface to move longitudinally (and hence in rotational phase). The backwards drift seen in \gaia\ in Fig.~\ref{fig:gaia_2d_lc} is also observed in TESS observations of CD~Ind when using the proposed frequency identifications of \citet{littlefield} and in SDSS~J084617.11+245344.1 \citep{littlefield_J0846}; however, in Paloma, the spin pulse seems to move forward in phase across the beat cycle, although this trend is less pronounced than in the aforementioned APs \citep{littlefield_J0846}. In a follow-up study, we intend to undertake a more comprehensive analysis of the motion of the spin pulse in the asynchronous mCVs observed by TESS and Kepler K2, with a focus on understanding this movement using the framework of \citet{gs97} and using it to potentially test the \citet{wang2020} AP frequency identifications.

    Fig.~\ref{fig:gaia_2d_lc} highlights the consequences of the selection of $2\omega-\Omega$ compared to $\omega$. Our identification of $\omega$ is consistent with there being accretion regions on opposite hemispheres of the WD, each active for alternating halves of the beat cycle based on the presence of a $\sim180^{\circ}$ phase jump. Conversely, $2\omega-\Omega$ remains relatively stable in phase over the beat cycle. Had we identified the 19~\cpd\ signal as $\omega$ instead of $2\omega-\Omega$, it would have appeared as though the WD has two accretion regions in the same hemisphere. The identification of $\omega$ therefore has significant implications for the accretion geometry. It is not an issue limited to \gaia; in their analyses of the TESS light curve of CD~Ind, \citet{littlefield_cd_ind} and \citet{hakala_cd_ind} reached differing interpretations of the accretion geometry depending on the identification of $\omega$ vs. $2\omega-\Omega$; see also \citet{mason20}, who compares the periodicities seen in CD Ind with those of BY Cam.


    As with CD~Ind, we argue that the $2\omega-\Omega$ sideband is the largest-amplitude frequency in the power spectrum, a result that was justified mathematically by \citet{wynn92} for IPs. \citet{littlefield_cd_ind}  argued, on the basis of a series of papers \citep{mason89,mason95, mason98, silber92, silber97, zucker95}, that $2\omega-\Omega$ can be the dominant signal in a Lomb-Scargle power spectrum of an AP that undergoes pole switching during an observation. \citet{littlefield_cd_ind} pointed out that the Lomb-Scargle algorithm does not contemplate the possibility that a signal can undergo phase shifts. On the contrary, when presented with a signal that undergoes abrupt phase shifts, it rewards frequency identifications that keep a phase-shifting signal in phase. This is why the non-intuitive $2\omega-\Omega$ sideband can dominate a power spectrum even if on short timescales the light curve is modulated at $\omega$ \citep{mason20}.

    Despite its ambiguous orbital period, all plausible identifications of $\Omega$ would result in \gaia\ having the second-shortest orbital period of any polar (synchronous or asynchronous). Only Gaia~19bxc, which is believed to have a Population II donor \citep{Galiullin}, has a shorter orbital period (64.4~min) than \gaia\ (77.55~min).

    \subsection{All other available data are consistent with an mCV classification for \gaia}

    Both the short-term variability in TESS and the long-term light curve from ATLAS support an mCV classification for \gaia. But since the system has not been previously classified in the literature, it is prudent to review other evidence relevant to its proper classification.

    \citet{abrahams} studied the Gaia $B_p -R_p$ color indices and absolute $G$ magnitudes of a variety of CV subtypes, including polars. The absolute Gaia DR3 $G$ magnitude of \gaia\ is 10.5, while its $B_p -R_p$ color index is 0.64. The color-magnitude diagrams in Fig.~3 of \citet{abrahams} show that these values are consistent with both dwarf novae and magnetic CVs, while their Fig.~1 predicts that such a CV will have an orbital period below the period gap. The system's absolute magnitude and color index are therefore consistent with our classification of it as a short-period mCV.

    Furthermore, mCVs generally have X-ray counterparts, and consistent with that trend,  \gaia\ is 34" from a known X-ray source, 1RXS J232928.0-161654, which is more than twice the positional error for the X-ray source (16\arcsec). A subsequent Swift XRT observation of \gaia\ on 2025 September 17 (obsid 03000098001) confirmed that \gaia\ has an X-ray counterpart just 0.24\arcsec\ away, with a positional uncertainty of 2.6\arcsec, based on astrometry from the University of Leicester's ``Build Swift-XRT products'' page.\footnote{\url{https://www.swift.ac.uk/user_objects/}}

    While the available evidence are consistent with \gaia\ being an mCV, spectroscopic follow-up is necessary to confirm this proposed classification.

\section{\ztf: an eclipsing mCV with a large spin-to-orbit ratio}

    \subsection{Overview}

    \ztf\ has received no significant attention in the literature except for \citet{takata22}, who presented its TESS light curve and power spectrum in their Figure~14. In that study, it is referred to as Gaia~DR2~4321588332240659584 (G432 in their shorthand), and its other identifiers include TIC~1919787634, IPHAS~J192530.55+155426.5, and DDE~182. Its Gaia eDR3 distance is $504^{+24}_{-27}$~pc \citep{BJ21}.

    The primary notability of \ztf\ in \citet{takata22} is that the authors detected two fundamental frequencies in the TESS power spectrum, which they denoted as $F_0 = 18.5521$~\cpd\ and $F_u = 32.17$~\cpd. They further identified both the $F_u/2$ subharmonic and the $2F_0$ harmonic, speculated that the object might be an IP, but pointed out that there was a risk that some of the variability might be attributable to a blended source.

    \ztf\ has G = 9.6 and Bp-Rp = 0.81. \citet{abrahams} show that this color-magnitude combination is typical of mCVs under the period gap. It has an X-ray counterpart \citep[2SXPS J192530.4+155424;][]{takata22} and has been identified as a source with excess H$\alpha$ emission in the INT/WFC Photometric H$\alpha$ Survey \citep[IPHAS;][]{witham}, both of which are consistent with the system being an mCV.

\subsection{ATLAS and ZTF data rule out blending}

    \begin{figure*}
        \includegraphics[width=\textwidth]{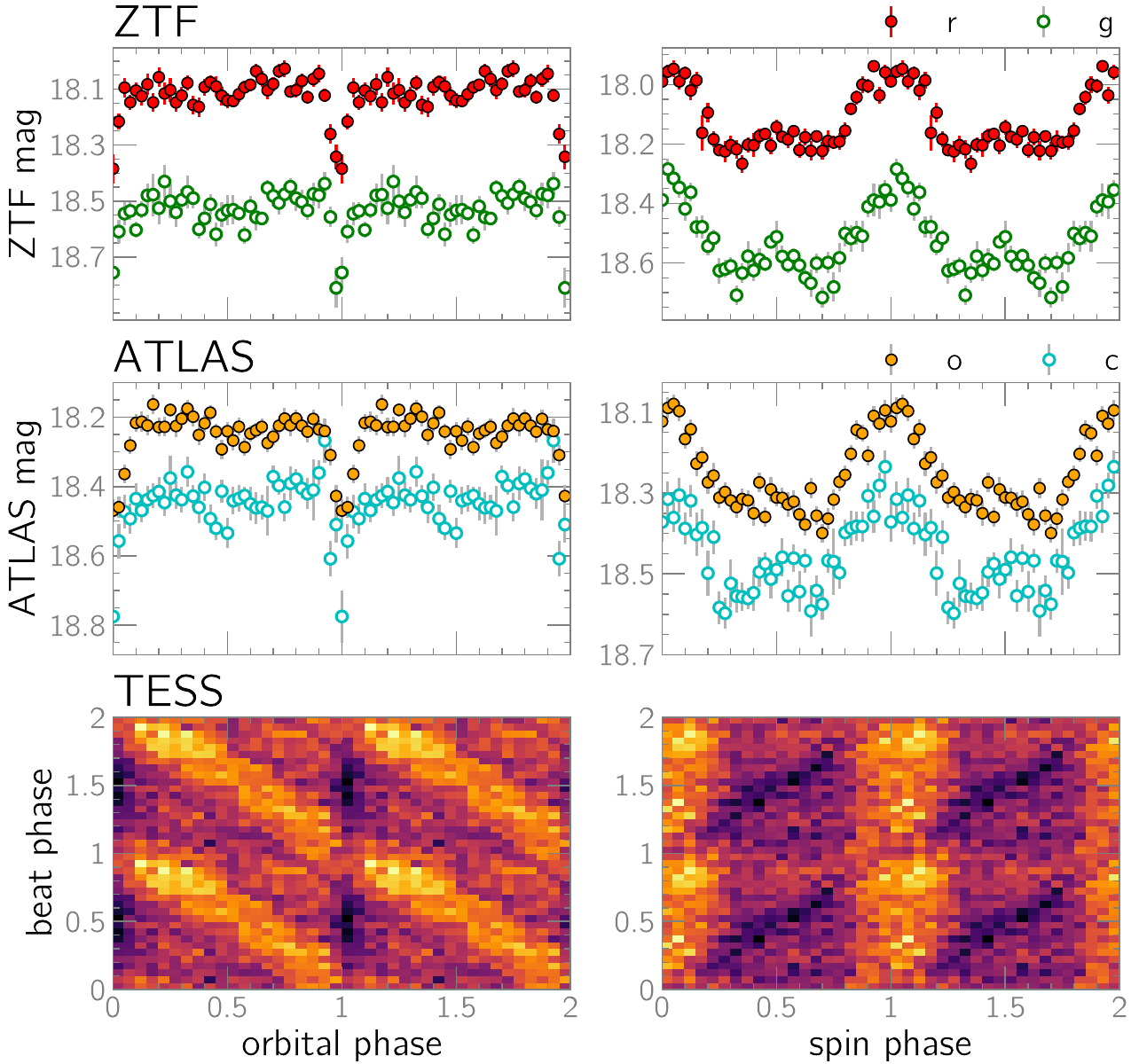}
        \caption{Phased 1D light curves of \ztf\ using public multiband photometry from ZTF (top row) and ATLAS (middle row). The bottom row shows 2D light curves from TESS Sectors 40, 51, and 81. All datasets show an eclipse when phased to the previously unidentified 16.0851~\cpd\ signal, whose presence in the higher-angular-resolution ZTF and ATLAS data verifies that it is not caused by blending in the TESS data. Notably, the phase-averaged spin profile looks deceptively similar to that observed in some synchronous polars, and it does not show the abrupt phase jumps of APs.
        \label{fig:3_light_curves_of_ztf18}}
    \end{figure*}

    \citet{takata22} cautioned that the apparent presence of multiple periodicities in \ztf\ might be caused other blended sources, which is a valid concern given the very coarse pixel scale of TESS (21\arcsec\ px$^{-1}$). This is therefore the starting point of our analysis.

    The ATLAS and ZTF photometry was obtained at significantly higher angular resolutions than TESS and can be used to investigate the possibility of blending. The ATLAS data have a typical point-spread function\footnote{\url{https://atlas.fallingstar.com/specifications.php}} (PSF) of $<$4~arcsec, while ZTF has an expected PSF of $\sim2$~arcsec. If the multiple periodicities in TESS were caused by blending with a background source, they would not be present in the higher-resolution ATLAS and ZTF data.

    Using these supplemental data, we confirm that the multiple periodicities of \ztf\ are intrinsic to the source and are not produced by blending. In Fig.~\ref{fig:3_light_curves_of_ztf18}, we phase both datasets to the 16.0851~\cpd\ and 18.55~\cpd\ frequencies from \citet{takata22}. Notably, the phased ATLAS and ZTF data elucidate the nature of the mysterious 16.0851~\cpd\ signal: previously identified as a subharmonic in \citet{takata22}, it is actually the binary orbital frequency based on the presence of grazing ($\sim0.3$ mag) eclipses.

    We conclude that the two frequencies in the TESS data from \citet{takata22} cannot be attributed to blending and that \ztf\ is indeed multiperiodic showing shallow eclipses at its 89.5~min orbital period.

    \begin{figure*}
        \includegraphics[width=\textwidth]{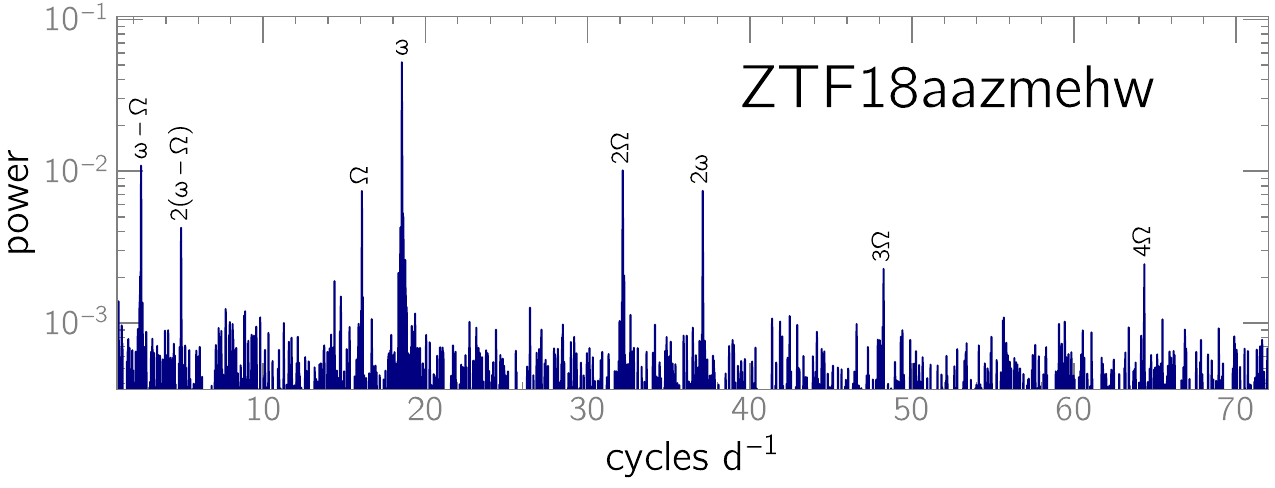}
        \caption{TESS power spectrum of \ztf\ in sectors 40, 54, and 81. $\Omega$ and $\omega$ denote the binary orbital and WD spin frequencies, respectively.     \label{fig:2SXPS_power}     }
    \end{figure*}

\subsection{TESS data}

    The TESS observations of \ztf\ were obtained in the full-frame mode during sectors 14, 40, 54, and 81 at a variety of cadences (30~min in sector 14, 10~min in sectors 40 and 54, and 200~s in sector 81). We limit our analysis to sectors 40, 54, and 81 because of their superior sampling cadence.

    The power spectrum of the TESS data (Fig.~\ref{fig:2SXPS_power})
    shows two fundamental frequencies: $\Omega =  16.0851$~\cpd\ (89.5~min) and the presumptive spin frequency $\omega = 18.55217$~\cpd\ (77.6~min). Moreover, there is a sideband frequency at 2.467~\cpd\ (9.73~h), equivalent to the beat frequency $\omega-\Omega$.

    Phasing the TESS, ATLAS, and ZTF data to an orbital ephemeris of \begin{equation} T_{min}[BJD] = 2459457.873(2) + 0.0621693(2) \times E\end{equation} shows the eclipse at the same orbital phase in all three datasets. Repeating this procedure for the presumptive spin frequency, we find that this signal in the TESS dataset has an identical counterpart in the ZTF and ATLAS data. The second column in Fig.~\ref{fig:3_light_curves_of_ztf18} phases this signal to an ephemeris of \begin{equation}
        T_{max}[BJD] = 2459457.837(2) + 0.0539020(2) \times E
    \end{equation} and shows that the phase-averaged profile is consistent with the one shown in \citet{takata22}.

\begin{figure}
            \includegraphics[width=\columnwidth]{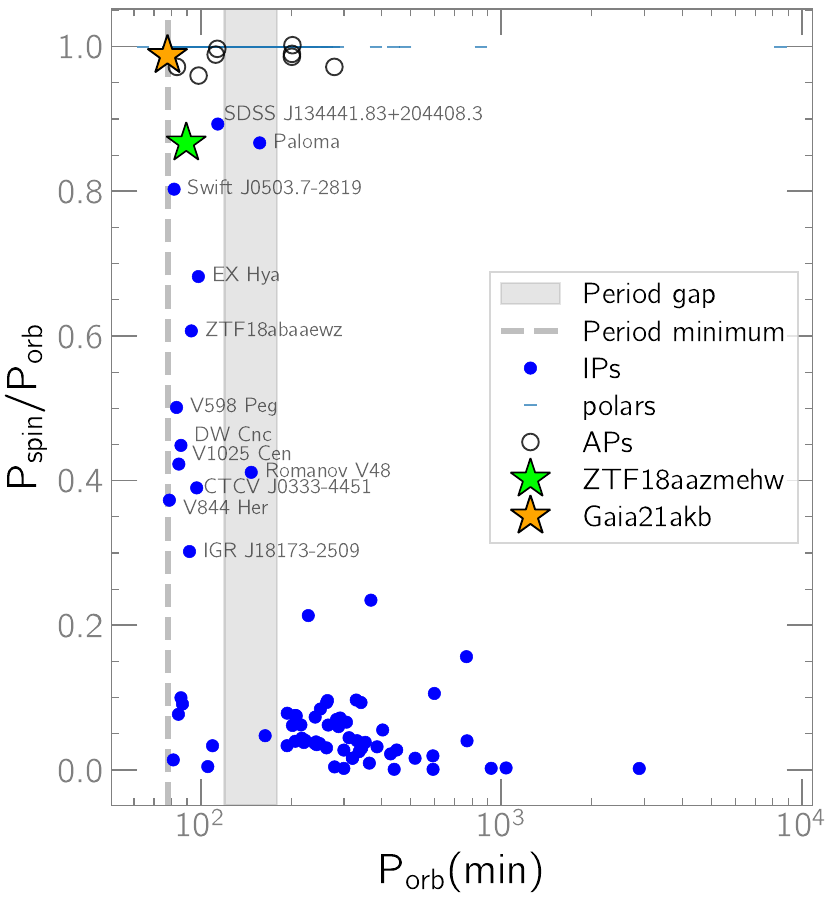}
            \caption{Asynchronous mCVs, with \ratio\ sourced primarily from Koji Mukai's online catalog. The new systems reported in this study are shown with colored stars for improved visibility. Above the period gap, asynchronous mCVs are heavily clustered below \ratio$<0.1$, and there are no known mCVs with $0.25\lesssim$\ratio$\lesssim 0.95$.  Conversely, this parameter space is well-populated in mCVs below the period gap, where \ratio\ is almost uniformly distributed between 0 and 1. For Swift J0503, we use \ratio\ from \citet{rawat22}.{\bf  We include V844~Her based on the photometric and spectroscopic analysis in Greiveldinger et al. (ApJ submitted). The parameters for CTCV~J0333-4451 are taken from \citet{J0333}.}
            \label{fig:mcvs}
            }

        \end{figure}

    With the orbital period having been identified by virtue of the eclipses, the 18.55217~\cpd\ signal is almost certainly the WD spin frequency. Both the amplitude and the shape of the phased signal (our Fig.~\ref{fig:3_light_curves_of_ztf18} and Fig.~14 in \citealt{takata22}) are consistent with the rotational profile of a polar. Furthermore, no other identification would make sense; in particular, if this were the $2\omega-\Omega$ sideband, it would mean that neither $\omega$ nor any of its harmonics are detectable. Our identifications of $\omega$ and $\Omega$ result in a spin-to-orbit ratio of 0.867, which is consistent with the continuum of spin-to-orbit ratios observed near the CV period minimum (Fig.~\ref{fig:mcvs}).

    However, \ztf\ shows a number of surprising differences from other such systems. Compared to many other well-observed mCVs with large \ratio, its power spectrum is startling in its simplicity: it contains just $\omega$, $\Omega$, and $\omega-\Omega$ and their harmonics without the usual forest of obscure sidebands. For comparison, we consider three mCVs with large \ratio: Paloma \citep[\ratio=0.87;][]{littlefield_J0846}, J1344 \cite[\ratio=0.893;][]{littlefield_J1344}, and Gaia18cja \citep[\ratio=0.607;][who refer to this object as 1631+69]{szkody24}. All three systems show exhibit complex behavior across their beat cycles, resulting in a litany of sidebands in their TESS power spectra as well as phase-averaged profiles that contain large phase jumps over the beat cycle. For example, Paloma's spin profile alternates between being single- and double-humped across the system's beat period. In contrast, this complex behavior is altogether absent in \ztf. Notably, the phase of the spin pulse appears constant across the beat cycle; there is no evidence of pole switching, and the spin profile shows very little variation across the beat cycle.

    Swift~J0503.7-2819 \citep[hereafter Swift~J0503;][]{halpern22, rawat22} is perhaps a more apt point of comparison for \ztf. Depending on the frequency identifications, \ratio\ is either 0.79 or 0.89, with \citet{rawat22} favoring the former. \citet{halpern22} found that it was unclear whether pole switching occurs in Swift~J0503, even though either value of \ratio\ is far above the maximum at which an accretion disk would be expected to exist if the system is in rotational equilibrium. \citet{halpern22} pointed out that the power spectrum of Swift~J0503 could be explained by the stream-overflow accretion model, wherein the WD accretes from both a disk-like structure and a portion of the accretion stream that overflows the rim and directly impacts the magnetosphere.

    On one hand, the absence of discernible pole switching in \ztf\ might be the result of the WD's magnetic axis being more closely aligned with its rotational axis than in other asynchronous mCVs. In such a scenario, the geometry of the accreting field lines would not change across the beat cycle, suppressing the extravagant variations that other asynchronous mCVs often show. But on the other hand, if the spin and magnetic axes were closely aligned, it would suppress the amplitude of $\omega$, contradicting the observations.

    \section{Non-detection of nova shells around \ztf and \gaia}

        Motivated by the theory that the asynchronous rotation in APs and some high-\ratio\ IPs might be caused by nova eruptions in a formerly synchronous system, \citet{pagnotta16} searched unsuccessfully for nova shells around four systems listed in Table~\ref{table:APs}: V1432~Aql, BY~Cam, CD~Ind, and EX~Hya. Similarly, \citet{sd22} reported non-detections of nova shells around Paloma (using its alias of RXJ0524+4244), BY~Cam, V1432~Aql, and V1500~Cyg. We therefore decided to search for any evidence of a nova shell around either \ztf\ or \gaia.

        \subsection{H$\alpha$ images of \ztf\ from the MDW Survey}

        \ztf\ has been imaged by the Mittelman-di Cicco-Walker (MDW) H$\alpha$ Sky Survey \citep[hereafter, MDW Survey;][]{mdw_survey_dr0, mdw_survey_dr1} and the aforementioned IPHAS. Founded and operated by amateur astronomers in a partnership with Columbia University, the ongoing MDW Survey will observe the entire sky with a pixel scale of 3.2\arcsec\ and a typical point-spread function of 6\arcsec\ using a 130mm f/4.5 apochromatic refractor. Its DR1 covers the sky north of the celestial equator.

        As an initial test of the MDW Survey's ability to recover faint nova shells, we inspected the MDW image of V1315~Aql, around which \citet{sahman15, sahman18} identified a faint nova shell. Although the nova shell in the MDW data was not as conspicuous as in \citet{sahman15, sahman18}, its general structure was discernible, giving us confidence that the MDW Survey has sufficient depth to meaningfully probe for nova shells.

        The MDW Survey's DR1 observed \ztf\, which we downloaded and searched for any evidence of a nova remnant. This image (field 0646 in the survey's nomenclature) consists of eight 20-min exposures, for a total exposure time of 160~min, obtained through an H$\alpha$ filter with a 3~nm bandpass. There is no evidence of a nova shell (Fig.~\ref{fig:nova_shell_images}).

        We also downloaded the IPHAS H$\alpha$ and $r$ images of \ztf\ with the objective of searching for a nova shell, but we found none in either the summed H$\alpha$ images or the scaled H$\alpha - r$ difference image. However, as \citet{sd22} pointed out, IPHAS images tend to have too short of an exposure time to detect faint nova shells, especially with the bright H$\alpha$ background often observed along the Galactic plane. Indeed, the images that we reviewed had exposure times of just 120~s.

        \subsection{LBT image of \gaia}

\begin{figure*}
           \plottwo{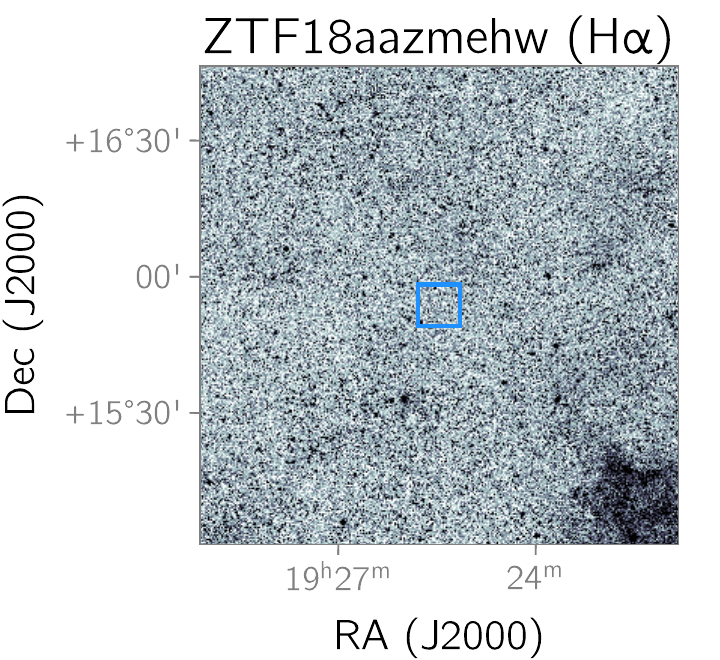}{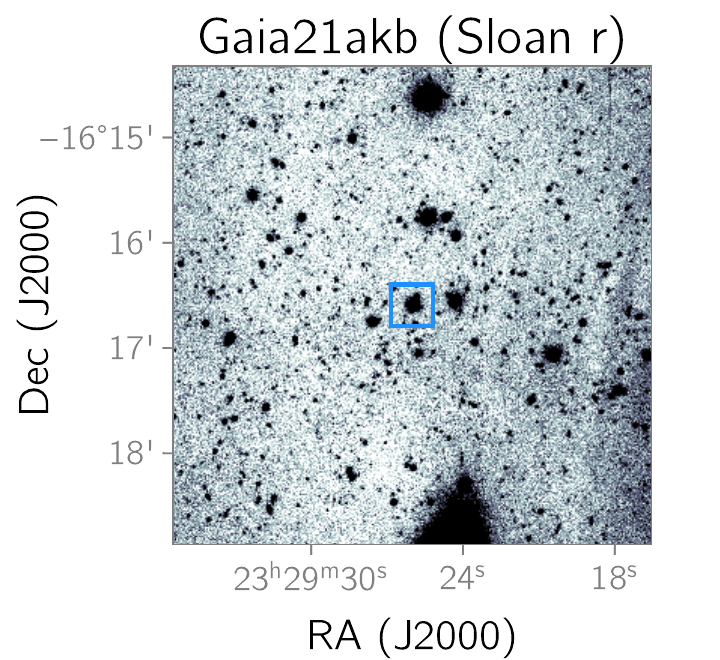}
            \caption{{\bf Left:} H$\alpha$ image of \ztf\ obtained by the MDW Survey.
            {\bf Right:} Broadband $r$-band image of \gaia, obtained with the LBC imager on the LBT. The extended source at the bottom of the frame is the galaxy IC~1491. In both panels, the position of the target is marked. Neither image shows any apparent evidence of a nova shell.\label{fig:nova_shell_images}}

        \end{figure*}

        In a similar spirit, we obtained a deep image of \gaia\ using the LBC Red camera \citep{lbc_giallongo, lbc_speziali} on the Large Binocular Telescope (LBT) on 2025 October 30. Unfortunately, no narrowband emission-line filters were available, so we used a Sloan $r$ filter. We obtained eight 250-sec exposures and coadded them after first subtracting bias-correcting and flat-fielding them. Shown in Fig.~\ref{fig:nova_shell_images}, this image has a limiting magnitude of 25.6 (for a signal-to-noise ratio of 5) and does not give any indication of a nova shell, but a deeper search using emission-line filters would offer a more decisive answer.

\section{Discussion and Conclusion}

    The identification of these two asynchronous mCVs expands the recently identified population of asynchronous mCVs below the period gap. Above the period gap, asynchronous mCVs overwhelmingly show  \ratio$\lesssim0.1 $, while below the gap, the distribution of \ratio\ forms a continuum without any obvious central tendency.

    An intriguing interpretation of this subset of mCVs is that they are evolving into synchronous polars for the first time. \citet{cr84} pointed out that angular-momentum losses in short-period asynchronous mCVs can cause the orbital separation to shrink within the WD's magnetospheric radius. There is a very clear prediction in the \citet{cr84} model: that some IPs rotationally synchronize into polars below the period gap. Thus, if the synchronization timescales are sufficiently long, we would expect to observe some IPs under the gap to show large \ratio\ values as they synchronize, as is now observed in Figure \ref{fig:mcvs}.

    One challenge with the \citet{cr84} model has been that the synchronization timescale is expected to be so short ($<<$ 10~Myr) that detecting an IP in the process of synchronizing would be improbable. Addressing this issue, \citet{littlefield_J1344} contended that the short theoretical timescale for the dynamic synchronization of mCVs is lengthened by an order of magnitude if the WD is massive. In this hypothesis, the transformation of the below-gap IPs with higher magnetic-field strength (B $\gtrsim$ 3 MG) into synchronized polars depends sensitively on the radius, and hence the mass, of the synchronizing WD.

    This scenario is an appealing explanation for the distribution in Fig.~\ref{fig:mcvs}, but it needs to be tested with (1) measurements of WD masses in systems with large \ratio\ and (2) measurements of their spin-period derivatives. In particular, all of the well studied APs have measured WD spin-period derivatives, directed towards synchronization on a time scale of 100 - 1000 years (e.g. \citealt{schmidt91, mason95, myers17}). A short-term disruption of synchronous rotation by a nova eruption is expected to be the cause of asynchronism in all quickly synchronizing APs. However, none of the below-gap IPs on the proposed IP-to-polar synchronization track have been observed long enough to establish their spin-period derivatives. For this reason, prolonged photometric monitoring of asynchronous mCVs will play an important role in elucidating the nature of the high \ratio\ mCVs.

    Finally, we note that the apparent absence of pole switching in \ztf\ (and possibly in Swift J0503) has noteworthy implications for the detectability of other such systems. Although it might seem strange to complain that an mCV power spectrum is too simple, \ztf\ would likely appear to be an ultrashort-period synchronous polar in our data if its orbital inclination were just slightly lower. Eclipses occur for only a small range of orbital inclinations ($i \gtrsim78^{\circ}$), but the spin modulation (assuming that the spin and magnetic axes are misaligned) should be observable for a much wider range of orbital inclinations. If it were not for the eclipses in \ztf, it might not be apparent that $\omega$ and $\Omega$ are different, and in isolation, the shape, amplitude, and phase stability of the spin-phased light curve in Fig.~\ref{fig:3_light_curves_of_ztf18} could reasonably be interpreted as the light curve of a synchronous system. Consequently, even a diligent observer might measure the spin period and mistakenly identify it as the orbital period, and in the Appendix, we suggest that this might explain the apparent ultrashort orbital period of another mCV, Gaia19bxc.

    In contrast, the complex power spectra of asynchronous mCVs like Paloma, J1344, and ZTF18abaaewz \citep{szkody24} make it obvious that $\Omega \neq \omega$, even though it might be very difficult to convincingly identify $\Omega$ and $\omega$ individually.

    \acknowledgments

    We thank the anonymous referee for reviewing this manuscript and offering a number of helpful recommendations for improving it.

    After we posted a preprint of this manuscript, Vladislav Dodon contacted CL to call his attention to \citet{Galiullin}. CL is grateful to Dr. Dodon for alerting him to this oversight.

    KI is supported by the Polish National Science Centre (NCN) grant 2024/55/D/ST9/01713.

    The LBT is an international collaboration among institutions in the United States, Italy and Germany. LBT Corporation Members are: The University of Arizona on behalf of the Arizona Board of Regents; Istituto Nazionale di Astrofisica, Italy; LBT Beteiligungsgesellschaft, Germany, representing the Max-Planck Society, The Leibniz Institute for Astrophysics Potsdam, and Heidelberg University; The Ohio State University, representing OSU, University of Notre Dame, University of Minnesota and University of Virginia. Observations have benefited from the use of ALTA Center (\url{alta.arcetri.inaf.it}) forecasts performed with the Astro-Meso-Nh model. Initialization data of the ALTA automatic forecast system come from the General Circulation Model (HRES) of the European Centre for Medium Range Weather Forecasts.

    Funding for the MDW Survey Project has been provided by the Michele and David Mittelman Family Foundation. David R. Mittelman, Dennis di Cicco, and Sean Walker are founding members of the survey and made possible the acquisition and reduction of the data. The Columbia University Astronomy Department is responsible for the final data reduction, calibration, and dissemination of the survey data.

\bibliography{bib.bib}

@ARTICLE{mdw_survey_dr0,
       author = {{Aftab}, Noor and {Zhang}, Xunhe (Andrew) and {Mittelman}, David R. and {di Cicco}, Dennis and {Walker}, Sean and {Sliski}, David H. and {Homa}, Julia and {Holm-Hansen}, Colin and {Putman}, Mary and {Schiminovich}, David and {Henden}, Arne and {Walker}, Gary},
        title = "{The MDW H{\ensuremath{\alpha}} Sky Survey: Data Release 0}",
      journal = {\aj},
         year = 2024,
        month = dec,
       volume = {168},
       number = {6},
          eid = {260},
        pages = {260},
          doi = {10.3847/1538-3881/ad7c48},
archivePrefix = {arXiv},
       eprint = {2409.11481},
 primaryClass = {astro-ph.IM},
       adsurl = {https://ui.adsabs.harvard.edu/abs/2024AJ....168..260A}
}

@ARTICLE{mdw_survey_dr1,
       author = {{Aftab}, Noor and {Zhang}, Xunhe (Andrew) and {Walker}, Sean and {di Cicco}, Dennis and {Mittelman}, David R. and {Gupta}, Sanya and {Saydjari}, Andrew K. and {Putman}, Mary and {Schiminovich}, David},
        title = "{The MDW H{\ensuremath{\alpha}} Sky Survey: Data Release 1}",
      journal = {\aj},
         year = 2026,
        month = jan,
       volume = {171},
       number = {1},
          eid = {17},
        pages = {17},
          doi = {10.3847/1538-3881/ae17b4},
archivePrefix = {arXiv},
       eprint = {2510.22900},
 primaryClass = {astro-ph.IM},
       adsurl = {https://ui.adsabs.harvard.edu/abs/2026AJ....171...17A}
}

@ARTICLE{abrahams,
       author = {{Abrahams}, Ellianna S. and {Bloom}, Joshua S. and {Szkody}, Paula and {Rix}, Hans-Walter and {Mowlavi}, Nami},
        title = "{Informing the Cataclysmic Variable Sequence from Gaia Data: The Orbital-period-Color-Absolute-magnitude Relationship}",
      journal = {\apj},
         year = 2022,
        month = oct,
       volume = {938},
       number = {1},
          eid = {46},
        pages = {46},
          doi = {10.3847/1538-4357/ac87ab},
       adsurl = {https://ui.adsabs.harvard.edu/abs/2022ApJ...938...46A}
}

@ARTICLE{BJ21,
       author = {{Bailer-Jones}, C.~A.~L. and {Rybizki}, J. and {Fouesneau}, M. and {Demleitner}, M. and {Andrae}, R.},
        title = "{Estimating Distances from Parallaxes. V. Geometric and Photogeometric Distances to 1.47 Billion Stars in Gaia Early Data Release 3}",
      journal = {\aj},
         year = 2021,
        month = mar,
       volume = {161},
       number = {3},
          eid = {147},
        pages = {147},
          doi = {10.3847/1538-3881/abd806},
archivePrefix = {arXiv},
       eprint = {2012.05220},
 primaryClass = {astro-ph.SR},
       adsurl = {https://ui.adsabs.harvard.edu/abs/2021AJ....161..147B}
}

@ARTICLE{ztf_paper,
       author = {{Bellm}, Eric C. and {Kulkarni}, Shrinivas R. and {Graham}, Matthew J. and {Dekany}, Richard and {Smith}, Roger M. and {Riddle}, Reed and {Masci}, Frank J. and {Helou}, George and {Prince}, Thomas A. and {Adams}, Scott M. and {Barbarino}, C. and {Barlow}, Tom and {Bauer}, James and {Beck}, Ron and {Belicki}, Justin and {Biswas}, Rahul and {Blagorodnova}, Nadejda and {Bodewits}, Dennis and {Bolin}, Bryce and {Brinnel}, Valery and {Brooke}, Tim and {Bue}, Brian and {Bulla}, Mattia and {Burruss}, Rick and {Cenko}, S. Bradley and {Chang}, Chan-Kao and {Connolly}, Andrew and {Coughlin}, Michael and {Cromer}, John and {Cunningham}, Virginia and {De}, Kishalay and {Delacroix}, Alex and {Desai}, Vandana and {Duev}, Dmitry A. and {Eadie}, Gwendolyn and {Farnham}, Tony L. and {Feeney}, Michael and {Feindt}, Ulrich and {Flynn}, David and {Franckowiak}, Anna and {Frederick}, S. and {Fremling}, C. and {Gal-Yam}, Avishay and {Gezari}, Suvi and {Giomi}, Matteo and {Goldstein}, Daniel A. and {Golkhou}, V. Zach and {Goobar}, Ariel and {Groom}, Steven and {Hacopians}, Eugean and {Hale}, David and {Henning}, John and {Ho}, Anna Y.~Q. and {Hover}, David and {Howell}, Justin and {Hung}, Tiara and {Huppenkothen}, Daniela and {Imel}, David and {Ip}, Wing-Huen and {Ivezi{\'c}}, {\v{Z}}eljko and {Jackson}, Edward and {Jones}, Lynne and {Juric}, Mario and {Kasliwal}, Mansi M. and {Kaspi}, S. and {Kaye}, Stephen and {Kelley}, Michael S.~P. and {Kowalski}, Marek and {Kramer}, Emily and {Kupfer}, Thomas and {Landry}, Walter and {Laher}, Russ R. and {Lee}, Chien-De and {Lin}, Hsing Wen and {Lin}, Zhong-Yi and {Lunnan}, Ragnhild and {Giomi}, Matteo and {Mahabal}, Ashish and {Mao}, Peter and {Miller}, Adam A. and {Monkewitz}, Serge and {Murphy}, Patrick and {Ngeow}, Chow-Choong and {Nordin}, Jakob and {Nugent}, Peter and {Ofek}, Eran and {Patterson}, Maria T. and {Penprase}, Bryan and {Porter}, Michael and {Rauch}, Ludwig and {Rebbapragada}, Umaa and {Reiley}, Dan and {Rigault}, Mickael and {Rodriguez}, Hector and {van Roestel}, Jan and {Rusholme}, Ben and {van Santen}, Jakob and {Schulze}, S. and {Shupe}, David L. and {Singer}, Leo P. and {Soumagnac}, Maayane T. and {Stein}, Robert and {Surace}, Jason and {Sollerman}, Jesper and {Szkody}, Paula and {Taddia}, F. and {Terek}, Scott and {Van Sistine}, Angela and {van Velzen}, Sjoert and {Vestrand}, W. Thomas and {Walters}, Richard and {Ward}, Charlotte and {Ye}, Quan-Zhi and {Yu}, Po-Chieh and {Yan}, Lin and {Zolkower}, Jeffry},
        title = "{The Zwicky Transient Facility: System Overview, Performance, and First Results}",
      journal = {\pasp},
         year = 2019,
        month = jan,
       volume = {131},
       number = {995},
        pages = {018002},
          doi = {10.1088/1538-3873/aaecbe},
archivePrefix = {arXiv},
       eprint = {1902.01932},
 primaryClass = {astro-ph.IM},
       adsurl = {https://ui.adsabs.harvard.edu/abs/2019PASP..131a8002B}
}

@ARTICLE{cr84,
       author = {{Chanmugam}, G. and {Ray}, A.},
        title = "{The rotational and orbital evolution of cataclysmic binaries containing magnetic white dwarfs.}",
      journal = {\apj},
         year = 1984,
        month = oct,
       volume = {285},
        pages = {252-257},
          doi = {10.1086/162499},
       adsurl = {https://ui.adsabs.harvard.edu/abs/1984ApJ...285..252C}
}

@ARTICLE{duffy22,
       author = {{Duffy}, C. and {Ramsay}, G. and {Wu}, Kinwah and {Mason}, Paul A. and {Hakala}, P. and {Steeghs}, D. and {Wood}, M.~A.},
        title = "{Short-duration accretion states of Polars as seen in TESS and ZTF data}",
      journal = {\mnras},
         year = 2022,
        month = nov,
       volume = {516},
       number = {3},
        pages = {3144-3158},
          doi = {10.1093/mnras/stac2482},
archivePrefix = {arXiv},
       eprint = {2208.14855},
 primaryClass = {astro-ph.SR},
       adsurl = {https://ui.adsabs.harvard.edu/abs/2022MNRAS.516.3144D}
}

@ARTICLE{halpern22,
       author = {{Halpern}, J.~P.},
        title = "{Swift J0503.7-2819: a Short-period Asynchronous Polar or Stream-fed Intermediate Polar}",
      journal = {\apj},
         year = 2022,
        month = aug,
       volume = {934},
       number = {2},
          eid = {123},
        pages = {123},
          doi = {10.3847/1538-4357/ac7d50},
archivePrefix = {arXiv},
       eprint = {2206.14770},
 primaryClass = {astro-ph.HE},
       adsurl = {https://ui.adsabs.harvard.edu/abs/2022ApJ...934..123H}
}

@ARTICLE{hakala_cd_ind,
       author = {{Hakala}, Pasi and {Ramsay}, Gavin and {Potter}, Stephen B. and {Beardmore}, Andrew and {Buckley}, David A.~H. and {Wynn}, Graham},
        title = "{TESS observations of the asynchronous polar CD Ind: mapping the changing accretion geometry}",
      journal = {\mnras},
         year = 2019,
        month = jun,
       volume = {486},
       number = {2},
        pages = {2549-2556},
          doi = {10.1093/mnras/stz992},
archivePrefix = {arXiv},
       eprint = {1904.02949},
 primaryClass = {astro-ph.SR},
       adsurl = {https://ui.adsabs.harvard.edu/abs/2019MNRAS.486.2549H}
}

@ARTICLE{Galiullin,
       author = {{Galiullin}, Ilkham and {Rodriguez}, Antonio C. and {El-Badry}, Kareem and {Caiazzo}, Ilaria and {Szkody}, Paula and {Nagarajan}, Pranav and {Whitebook}, Samuel},
        title = "{Optical Spectroscopy of the Most Compact Accreting Binary Harboring a Magnetic White Dwarf and a Hydrogen-rich Donor}",
      journal = {\apjl},
         year = 2025,
        month = sep,
       volume = {990},
       number = {2},
          eid = {L57},
        pages = {L57},
          doi = {10.3847/2041-8213/adff82},
archivePrefix = {arXiv},
       eprint = {2508.20170},
 primaryClass = {astro-ph.SR},
       adsurl = {https://ui.adsabs.harvard.edu/abs/2025ApJ...990L..57G}
}

@ARTICLE{edr3,
       author = {{Gaia Collaboration} and {Brown}, A.~G.~A. and {Vallenari}, A. and {Prusti}, T. and {de Bruijne}, J.~H.~J. and {Babusiaux}, C. and {Biermann}, M. and {Creevey}, O.~L. and {Evans}, D.~W. and {Eyer}, L. and {Hutton}, A. and {Jansen}, F. and {Jordi}, C. and {Klioner}, S.~A. and {Lammers}, U. and {Lindegren}, L. and {Luri}, X. and {Mignard}, F. and {Panem}, C. and {Pourbaix}, D. and {Randich}, S. and {Sartoretti}, P. and {Soubiran}, C. and {Walton}, N.~A. and {Arenou}, F. and {Bailer-Jones}, C.~A.~L. and {Bastian}, U. and {Cropper}, M. and {Drimmel}, R. and {Katz}, D. and {Lattanzi}, M.~G. and {van Leeuwen}, F. and {Bakker}, J. and {Cacciari}, C. and {Casta{\~n}eda}, J. and {De Angeli}, F. and {Ducourant}, C. and {Fabricius}, C. and {Fouesneau}, M. and {Fr{\'e}mat}, Y. and {Guerra}, R. and {Guerrier}, A. and {Guiraud}, J. and {Jean-Antoine Piccolo}, A. and {Masana}, E. and {Messineo}, R. and {Mowlavi}, N. and {Nicolas}, C. and {Nienartowicz}, K. and {Pailler}, F. and {Panuzzo}, P. and {Riclet}, F. and {Roux}, W. and {Seabroke}, G.~M. and {Sordo}, R. and {Tanga}, P. and {Th{\'e}venin}, F. and {Gracia-Abril}, G. and {Portell}, J. and {Teyssier}, D. and {Altmann}, M. and {Andrae}, R. and {Bellas-Velidis}, I. and {Benson}, K. and {Berthier}, J. and {Blomme}, R. and {Brugaletta}, E. and {Burgess}, P.~W. and {Busso}, G. and {Carry}, B. and {Cellino}, A. and {Cheek}, N. and {Clementini}, G. and {Damerdji}, Y. and {Davidson}, M. and {Delchambre}, L. and {Dell'Oro}, A. and {Fern{\'a}ndez-Hern{\'a}ndez}, J. and {Galluccio}, L. and {Garc{\'\i}a-Lario}, P. and {Garcia-Reinaldos}, M. and {Gonz{\'a}lez-N{\'u}{\~n}ez}, J. and {Gosset}, E. and {Haigron}, R. and {Halbwachs}, J. -L. and {Hambly}, N.~C. and {Harrison}, D.~L. and {Hatzidimitriou}, D. and {Heiter}, U. and {Hern{\'a}ndez}, J. and {Hestroffer}, D. and {Hodgkin}, S.~T. and {Holl}, B. and {Jan{\ss}en}, K. and {Jevardat de Fombelle}, G. and {Jordan}, S. and {Krone-Martins}, A. and {Lanzafame}, A.~C. and {L{\"o}ffler}, W. and {Lorca}, A. and {Manteiga}, M. and {Marchal}, O. and {Marrese}, P.~M. and {Moitinho}, A. and {Mora}, A. and {Muinonen}, K. and {Osborne}, P. and {Pancino}, E. and {Pauwels}, T. and {Petit}, J. -M. and {Recio-Blanco}, A. and {Richards}, P.~J. and {Riello}, M. and {Rimoldini}, L. and {Robin}, A.~C. and {Roegiers}, T. and {Rybizki}, J. and {Sarro}, L.~M. and {Siopis}, C. and {Smith}, M. and {Sozzetti}, A. and {Ulla}, A. and {Utrilla}, E. and {van Leeuwen}, M. and {van Reeven}, W. and {Abbas}, U. and {Abreu Aramburu}, A. and {Accart}, S. and {Aerts}, C. and {Aguado}, J.~J. and {Ajaj}, M. and {Altavilla}, G. and {{\'A}lvarez}, M.~A. and {{\'A}lvarez Cid-Fuentes}, J. and {Alves}, J. and {Anderson}, R.~I. and {Anglada Varela}, E. and {Antoja}, T. and {Audard}, M. and {Baines}, D. and {Baker}, S.~G. and {Balaguer-N{\'u}{\~n}ez}, L. and {Balbinot}, E. and {Balog}, Z. and {Barache}, C. and {Barbato}, D. and {Barros}, M. and {Barstow}, M.~A. and {Bartolom{\'e}}, S. and {Bassilana}, J. -L. and {Bauchet}, N. and {Baudesson-Stella}, A. and {Becciani}, U. and {Bellazzini}, M. and {Bernet}, M. and {Bertone}, S. and {Bianchi}, L. and {Blanco-Cuaresma}, S. and {Boch}, T. and {Bombrun}, A. and {Bossini}, D. and {Bouquillon}, S. and {Bragaglia}, A. and {Bramante}, L. and {Breedt}, E. and {Bressan}, A. and {Brouillet}, N. and {Bucciarelli}, B. and {Burlacu}, A. and {Busonero}, D. and {Butkevich}, A.~G. and {Buzzi}, R. and {Caffau}, E. and {Cancelliere}, R. and {C{\'a}novas}, H. and {Cantat-Gaudin}, T. and {Carballo}, R. and {Carlucci}, T. and {Carnerero}, M.~I. and {Carrasco}, J.~M. and {Casamiquela}, L. and {Castellani}, M. and {Castro-Ginard}, A. and {Castro Sampol}, P. and {Chaoul}, L. and {Charlot}, P. and {Chemin}, L. and {Chiavassa}, A. and {Cioni}, M. -R.~L. and {Comoretto}, G. and {Cooper}, W.~J. and {Cornez}, T. and {Cowell}, S. and {Crifo}, F. and {Crosta}, M. and {Crowley}, C. and {Dafonte}, C. and {Dapergolas}, A. and {David}, M. and {David}, P.},
        title = "{Gaia Early Data Release 3. Summary of the contents and survey properties}",
      journal = {\aap},
         year = 2021,
        month = may,
       volume = {649},
          eid = {A1},
        pages = {A1},
          doi = {10.1051/0004-6361/202039657},
archivePrefix = {arXiv},
       eprint = {2012.01533},
 primaryClass = {astro-ph.GA},
       adsurl = {https://ui.adsabs.harvard.edu/abs/2021A&A...649A...1G}
}

@ARTICLE{gs97,
       author = {{Geckeler}, R.~D. and {Staubert}, R.},
        title = "{Periodic changes of the accretion geometry in the nearly-synchronous polar RX J1940.1-1025.}",
      journal = {\aap},
         year = 1997,
        month = sep,
       volume = {325},
        pages = {1070-1076},
       adsurl = {https://ui.adsabs.harvard.edu/abs/1997A&A...325.1070G}
}

@ARTICLE{lbc_giallongo,
       author = {{Giallongo}, E. and {Ragazzoni}, R. and {Grazian}, A. and {Baruffolo}, A. and {Beccari}, G. and {de Santis}, C. and {Diolaiti}, E. and {di Paola}, A. and {Farinato}, J. and {Fontana}, A. and {Gallozzi}, S. and {Gasparo}, F. and {Gentile}, G. and {Green}, R. and {Hill}, J. and {Kuhn}, O. and {Pasian}, F. and {Pedichini}, F. and {Radovich}, M. and {Salinari}, P. and {Smareglia}, R. and {Speziali}, R. and {Testa}, V. and {Thompson}, D. and {Vernet}, E. and {Wagner}, R.~M.},
        title = "{The performance of the blue prime focus large binocular camera at the large binocular telescope}",
      journal = {\aap},
         year = 2008,
        month = apr,
       volume = {482},
       number = {1},
        pages = {349-357},
          doi = {10.1051/0004-6361:20078402},
archivePrefix = {arXiv},
       eprint = {0801.1474},
 primaryClass = {astro-ph},
       adsurl = {https://ui.adsabs.harvard.edu/abs/2008A&A...482..349G}
}

@ARTICLE{v844her_paper1,
       author = {{Greiveldinger}, Anousha and {Garnavich}, Peter and {Littlefield}, Colin and {Kennedy}, Mark R. and {Halpern}, Jules P. and {Thorstensen}, John R. and {Szkody}, Paula and {Oksanen}, Arto and {Boyle}, Rebecca S.},
        title = "{A Surprising Periodicity Detected during a Super-outburst of V844 Herculis by TESS}",
      journal = {\apj},
         year = 2023,
        month = oct,
       volume = {955},
       number = {2},
          eid = {150},
        pages = {150},
          doi = {10.3847/1538-4357/acf21b},
archivePrefix = {arXiv},
       eprint = {2308.10344},
 primaryClass = {astro-ph.SR},
       adsurl = {https://ui.adsabs.harvard.edu/abs/2023ApJ...955..150G}
}

@ARTICLE{halpern24,
       author = {{Halpern}, J.~P.},
        title = "{Resolving the Periods of the Asynchronous Polar 1RXS J083842.1{\textendash}282723}",
      journal = {\apj},
         year = 2024,
        month = mar,
       volume = {963},
       number = {2},
          eid = {78},
        pages = {78},
          doi = {10.3847/1538-4357/ad24ed},
       adsurl = {https://ui.adsabs.harvard.edu/abs/2024ApJ...963...78H}
}

@ARTICLE{howell_k2,
       author = {{Howell}, Steve B. and {Sobeck}, Charlie and {Haas}, Michael and {Still}, Martin and {Barclay}, Thomas and {Mullally}, Fergal and {Troeltzsch}, John and {Aigrain}, Suzanne and {Bryson}, Stephen T. and {Caldwell}, Doug and {Chaplin}, William J. and {Cochran}, William D. and {Huber}, Daniel and {Marcy}, Geoffrey W. and {Miglio}, Andrea and {Najita}, Joan R. and {Smith}, Marcie and {Twicken}, J.~D. and {Fortney}, Jonathan J.},
        title = "{The K2 Mission: Characterization and Early Results}",
      journal = {\pasp},
         year = 2014,
        month = apr,
       volume = {126},
       number = {938},
        pages = {398},
          doi = {10.1086/676406},
archivePrefix = {arXiv},
       eprint = {1402.5163},
 primaryClass = {astro-ph.IM},
       adsurl = {https://ui.adsabs.harvard.edu/abs/2014PASP..126..398H}
}

@ARTICLE{J0333,
       author = {{I{\l}kiewicz}, Krystian and {Scaringi}, Simone and {Veresvarska}, Martina and {De Martino}, Domitilla and {Littlefield}, Colin and {Knigge}, Christian and {Paice}, John A. and {Sahu}, Anwesha},
        title = "{Classifying Optical (Out)bursts in Cataclysmic Variables: The Distinct Observational Characteristics of Dwarf Novae, Micronovae, Stellar Flares, and Magnetic Gating}",
      journal = {\apjl},
         year = 2024,
        month = feb,
       volume = {962},
       number = {2},
          eid = {L34},
        pages = {L34},
          doi = {10.3847/2041-8213/ad243c},
archivePrefix = {arXiv},
       eprint = {2402.00553},
 primaryClass = {astro-ph.SR},
       adsurl = {https://ui.adsabs.harvard.edu/abs/2024ApJ...962L..34I}
}

@ARTICLE{gaia19bxc,
       author = {{Kato}, Taichi},
        title = "{Gaia19bxc: possible polar below the period minimum}",
      journal = {arXiv e-prints},
         year = 2022,
        month = apr,
          eid = {arXiv:2204.04603},
        pages = {arXiv:2204.04603},
          doi = {10.48550/arXiv.2204.04603},
archivePrefix = {arXiv},
       eprint = {2204.04603},
 primaryClass = {astro-ph.SR},
       adsurl = {https://ui.adsabs.harvard.edu/abs/2022arXiv220404603K}
}

@ARTICLE{Kato_V844her,
       author = {{Kato}, Taichi},
        title = "{Analysis of TESS observations of V844 Her during the 2020 superoutburst}",
      journal = {arXiv e-prints},
         year = 2022,
        month = may,
          eid = {arXiv:2205.05284},
        pages = {arXiv:2205.05284},
          doi = {10.48550/arXiv.2205.05284},
archivePrefix = {arXiv},
       eprint = {2205.05284},
 primaryClass = {astro-ph.SR},
       adsurl = {https://ui.adsabs.harvard.edu/abs/2022arXiv220505284K}
}

@ARTICLE{ex_hya,
       author = {{King}, A.~R. and {Wynn}, G.~A.},
        title = "{The spin period of EX Hydrae}",
      journal = {\mnras},
         year = 1999,
        month = nov,
       volume = {310},
       number = {1},
        pages = {203-209},
          doi = {10.1046/j.1365-8711.1999.02974.x},
archivePrefix = {arXiv},
       eprint = {astro-ph/9909312},
 primaryClass = {astro-ph},
       adsurl = {https://ui.adsabs.harvard.edu/abs/1999MNRAS.310..203K}
}

@ARTICLE{fr_lyn,
       author = {{Kolbin}, A.~I. and {Suslikov}, M.~V. and {Kochkina}, V. Yu. and {Borisov}, N.~V. and {Burenkov}, A.~N. and {Oparin}, D.~V.},
        title = "{SDSS J085414.02+390537.3{\textemdash}A New Asynchronous Polar}",
      journal = {Astronomy Letters},
         year = 2023,
        month = oct,
       volume = {49},
       number = {8},
        pages = {475-485},
          doi = {10.1134/S1063773723080029},
archivePrefix = {arXiv},
       eprint = {2308.04597},
 primaryClass = {astro-ph.HE},
       adsurl = {https://ui.adsabs.harvard.edu/abs/2023AstL...49..475K}
}

@ARTICLE{littlefield_J0846,
       author = {{Littlefield}, Colin and {Hoard}, D.~W. and {Garnavich}, Peter and {Szkody}, Paula and {Mason}, Paul A. and {Scaringi}, Simone and {Ilkiewicz}, Krystian and {Kennedy}, Mark R. and {Rappaport}, Saul A. and {Jayaraman}, Rahul},
        title = "{Kepler K2 and TESS Observations of Two Magnetic Cataclysmic Variables: The New Asynchronous Polar SDSS J084617.11+245344.1 and Paloma}",
      journal = {\aj},
         year = 2023,
        month = feb,
       volume = {165},
       number = {2},
          eid = {43},
        pages = {43},
          doi = {10.3847/1538-3881/aca1a5},
archivePrefix = {arXiv},
       eprint = {2205.02863},
 primaryClass = {astro-ph.SR},
       adsurl = {https://ui.adsabs.harvard.edu/abs/2023AJ....165...43L}
}

@ARTICLE{littlefield_cd_ind,
       author = {{Littlefield}, Colin and {Garnavich}, Peter and {Mukai}, Koji and {Mason}, Paul A. and {Szkody}, Paula and {Kennedy}, Mark and {Myers}, Gordon and {Schwarz}, Robert},
        title = "{Fast-cadence TESS Photometry and Doppler Tomography of the Asynchronous Polar CD Ind: A Revised Accretion Geometry from Newly Proposed Spin and Orbital Periods}",
      journal = {\apj},
         year = 2019,
        month = aug,
       volume = {881},
       number = {2},
          eid = {141},
        pages = {141},
          doi = {10.3847/1538-4357/ab2a17},
archivePrefix = {arXiv},
       eprint = {1903.00490},
 primaryClass = {astro-ph.SR},
       adsurl = {https://ui.adsabs.harvard.edu/abs/2019ApJ...881..141L}
}

@ARTICLE{littlefield15,
       author = {{Littlefield}, Colin and {Mukai}, Koji and {Mumme}, Raymond and
         {Cain}, Ryan and {Magno}, Katrina C. and {Corpuz}, Taylor and {Sand
        efur}, Davis and {Boyd}, David and {Cook}, Michael and
         {Ulowetz}, Joseph and {Martinez}, Luis},
        title = "{Periodic eclipse variations in asynchronous polar V1432 Aql: evidence of a shifting threading region}",
      journal = {\mnras},
         year = 2015,
        month = may,
       volume = {449},
       number = {3},
        pages = {3107-3120},
          doi = {10.1093/mnras/stv462},
       adsurl = {https://ui.adsabs.harvard.edu/abs/2015MNRAS.449.3107L}
}

@ARTICLE{littlefield,
       author = {{Littlefield}, Colin and {Garnavich}, Peter and {Mukai}, Koji and
         {Mason}, Paul A. and {Szkody}, Paula and {Kennedy}, Mark and
         {Myers}, Gordon and {Schwarz}, Robert},
        title = "{Fast-cadence TESS Photometry and Doppler Tomography of the Asynchronous Polar CD Ind: A Revised Accretion Geometry from Newly Proposed Spin and Orbital Periods}",
      journal = {\apj},
         year = 2019,
        month = aug,
       volume = {881},
       number = {2},
          eid = {141},
        pages = {141},
          doi = {10.3847/1538-4357/ab2a17},
archivePrefix = {arXiv},
       eprint = {1903.00490},
 primaryClass = {astro-ph.SR},
       adsurl = {https://ui.adsabs.harvard.edu/abs/2019ApJ...881..141L}
}

@ARTICLE{littlefield_J1344,
       author = {{Littlefield}, Colin and {Mason}, Paul A. and {Garnavich}, Peter and {Szkody}, Paula and {Thorstensen}, John and {Scaringi}, Simone and {I{\l}kiewicz}, Krystian and {Kennedy}, Mark R. and {Wells}, Natalie},
        title = "{SDSS J134441.83+204408.3: A Highly Asynchronous Short-period Magnetic Cataclysmic Variable with a 56 MG Field Strength}",
      journal = {\apjl},
         year = 2023,
        month = feb,
       volume = {943},
       number = {2},
          eid = {L24},
        pages = {L24},
          doi = {10.3847/2041-8213/acaf04},
archivePrefix = {arXiv},
       eprint = {2301.05723},
 primaryClass = {astro-ph.SR},
       adsurl = {https://ui.adsabs.harvard.edu/abs/2023ApJ...943L..24L}
}

@INPROCEEDINGS{mason95,
       author = {{Mason}, P.~A. and {Andronov}, I.~L. and {Kolesnikov}, S.~V. and {Pavlenko}, E.~P. and {Shakovskoy}, M.},
        title = "{Asynchronism and Multipole Accretion in BY Cam}",
    booktitle = {Magnetic Cataclysmic Variables},
         year = 1995,
       editor = {{Buckley}, D.~A.~H. and {Warner}, Brian},
       series = {Astronomical Society of the Pacific Conference Series},
       volume = {85},
        month = jan,
        pages = {496},
       adsurl = {https://ui.adsabs.harvard.edu/abs/1995ASPC...85..496M}
}

@ARTICLE{mason89,
       author = {{Mason}, Paul A. and {Liebert}, James and {Schmidt}, Gary D.},
        title = "{H0538+608: A Long-Period AM HER System Exhibiting Extreme Variations in Accretion Geometry}",
      journal = {\apj},
         year = 1989,
        month = nov,
       volume = {346},
        pages = {941},
          doi = {10.1086/168074},
       adsurl = {https://ui.adsabs.harvard.edu/abs/1989ApJ...346..941M}
}

@ARTICLE{mason98,
       author = {{Mason}, Paul A. and {Ramsay}, Gavin and {Andronov}, Ivan and {Kolesnikov}, Sergey and {Shakhovskoy}, Nickolay and {Pavlenko}, Elana},
        title = "{Evidence for pole switching in the magnetic cataclysmic variable BY Camelopardalis}",
      journal = {\mnras},
         year = 1998,
        month = apr,
       volume = {295},
       number = {3},
        pages = {511-518},
          doi = {10.1046/j.1365-8711.1998.01185.x},
       adsurl = {https://ui.adsabs.harvard.edu/abs/1998MNRAS.295..511M}
}

@ARTICLE{sd22,
       author = {{Sahman}, D.~I. and {Dhillon}, V.~S.},
        title = "{Searching for nova shells around cataclysmic variables - II. A second campaign}",
      journal = {\mnras},
         year = 2022,
        month = mar,
       volume = {510},
       number = {3},
        pages = {4180-4190},
          doi = {10.1093/mnras/stab3668},
archivePrefix = {arXiv},
       eprint = {2112.06629},
 primaryClass = {astro-ph.SR},
       adsurl = {https://ui.adsabs.harvard.edu/abs/2022MNRAS.510.4180S}
}

@ARTICLE{sahman15,
       author = {{Sahman}, D.~I. and {Dhillon}, V.~S. and {Knigge}, C. and {Marsh}, T.~R.},
        title = "{Searching for nova shells around cataclysmic variables}",
      journal = {\mnras},
         year = 2015,
        month = aug,
       volume = {451},
       number = {3},
        pages = {2863-2876},
          doi = {10.1093/mnras/stv1150},
archivePrefix = {arXiv},
       eprint = {1505.06048},
 primaryClass = {astro-ph.HE},
       adsurl = {https://ui.adsabs.harvard.edu/abs/2015MNRAS.451.2863S}
}

@ARTICLE{sahman18,
       author = {{Sahman}, D.~I. and {Dhillon}, V.~S. and {Littlefair}, S.~P. and {Hallinan}, G.},
        title = "{Discovery of an old nova shell surrounding the cataclysmic variable V1315 Aql}",
      journal = {\mnras},
         year = 2018,
        month = jul,
       volume = {477},
       number = {4},
        pages = {4483-4490},
          doi = {10.1093/mnras/sty950},
archivePrefix = {arXiv},
       eprint = {1804.05596},
 primaryClass = {astro-ph.HE},
       adsurl = {https://ui.adsabs.harvard.edu/abs/2018MNRAS.477.4483S}
}

@ARTICLE{schwarz_bycam,
       author = {{Schwarz}, R. and {Schwope}, A.~D. and {Staude}, A. and {Remillard}, R.~A.},
        title = "{Doppler tomography of the asynchronous polar BY Camelopardalis}",
      journal = {\aap},
         year = 2005,
        month = dec,
       volume = {444},
       number = {1},
        pages = {213-220},
          doi = {10.1051/0004-6361:20053711},
       adsurl = {https://ui.adsabs.harvard.edu/abs/2005A&A...444..213S}
}

@ARTICLE{schwarz_paloma,
       author = {{Schwarz}, R. and {Schwope}, A.~D. and {Staude}, A. and {Rau}, A. and {Hasinger}, G. and {Urrutia}, T. and {Motch}, C.},
        title = "{Paloma (RX J0524+42): the missing link in magnetic CV evolution?}",
      journal = {\aap},
         year = 2007,
        month = oct,
       volume = {473},
       number = {2},
        pages = {511-521},
          doi = {10.1051/0004-6361:20077684},
       adsurl = {https://ui.adsabs.harvard.edu/abs/2007A&A...473..511S}
}

@ARTICLE{silber92,
       author = {{Silber}, A. and {Bradt}, H.~V. and {Ishida}, M. and {Ohashi}, T. and {Remillard}, R.~A.},
        title = "{H0538+608 (= BY Camelopardalis): an Asynchronously Rotating AM Herculis Binary?}",
      journal = {\apj},
         year = 1992,
        month = apr,
       volume = {389},
        pages = {704},
          doi = {10.1086/171243},
       adsurl = {https://ui.adsabs.harvard.edu/abs/1992ApJ...389..704S}
}

@ARTICLE{silber97,
       author = {{Silber}, Andrew D. and {Szkody}, Paula and {Hoard}, D.~W. and {Hammergren}, M. and {Morgan}, J. and {Fierce}, E. and {Olsen}, K. and {Mason}, Paul A. and {Rolleston}, Robert and {Ruotsalainen}, Robert and {Pavlenko}, Elena P. and {Shakhovskoy}, Nickolay M. and {Shugarov}, Sergey and {Andronov}, Ivan L. and {Kolesnikov}, Sergey V. and {Naylor}, Tim and {Schmidt}, E.},
        title = "{The Noah Project: detection of the spin-orbit beat period of BYCamelopardalis}",
      journal = {\mnras},
         year = 1997,
        month = sep,
       volume = {290},
       number = {1},
        pages = {25-33},
          doi = {10.1093/mnras/290.1.25},
       adsurl = {https://ui.adsabs.harvard.edu/abs/1997MNRAS.290...25S}
}

@INPROCEEDINGS{mason15,
       author = {{Mason}, P.~A. and {Santana}, J.~B.},
        title = "{Low States of Polars: Catalina (CRTS) Light Curves}",
    booktitle = {The Golden Age of Cataclysmic Variables and Related Objects - III (Golden2015)},
         year = 2015,
        month = jan,
          eid = {16},
        pages = {16},
          doi = {10.22323/1.255.0016},
       adsurl = {https://ui.adsabs.harvard.edu/abs/2015gacv.workE..16M}
}

@ARTICLE{mason20,
       author = {{Mason}, Paul A. and {Morales}, John F. and {Littlefield}, Colin and {Garnavich}, Peter and {Pavlenko}, Elena P. and {Szkody}, Paula and {Kennedy}, Mark and {Myers}, Gordon and {Schwarz}, Robert and {Babina}, Julia V. and {Sosnovskij}, Alexsei A. and {Antonyuk}, Oksana I. and {Shugarov}, Sergey. Yu. and {Andreev}, Maksim V.},
        title = "{TESS photometry of the asynchronous polar CD Ind: A short period analog of BY Cam}",
      journal = {Advances in Space Research},
         year = 2020,
        month = sep,
       volume = {66},
       number = {5},
        pages = {1123-1138},
          doi = {10.1016/j.asr.2020.03.038},
       adsurl = {https://ui.adsabs.harvard.edu/abs/2020AdSpR..66.1123M}
}

@ARTICLE{mason22,
       author = {{Mason}, Paul A. and {Littlefield}, Colin and {Monroy}, Lorena C. and {Morales}, John F. and {Hakala}, Pasi and {Garnavich}, Peter and {Szkody}, Paula and {Kennedy}, Mark R. and {Ramsay}, Gavin and {Scaringi}, Simone},
        title = "{A Magnetic Valve at L1 Revealed in TESS Photometry of the Asynchronous Polar BY Cam}",
      journal = {\apj},
         year = 2022,
        month = oct,
       volume = {938},
       number = {2},
          eid = {142},
        pages = {142},
          doi = {10.3847/1538-4357/ac91cf},
archivePrefix = {arXiv},
       eprint = {2209.05524},
 primaryClass = {astro-ph.SR},
       adsurl = {https://ui.adsabs.harvard.edu/abs/2022ApJ...938..142M}
}

@ARTICLE{myers17,
       author = {{Myers}, Gordon and {Patterson}, Joseph and {de Miguel}, Enrique and {Hambsch}, Franz-Josef and {Monard}, Berto and {Bolt}, Greg and {McCormick}, Jennie and {Rea}, Robert and {Allen}, William},
        title = "{Resynchronization of the Asynchronous Polar CD Ind}",
      journal = {\pasp},
         year = 2017,
        month = apr,
       volume = {129},
       number = {974},
        pages = {044204},
          doi = {10.1088/1538-3873/aa54a8},
archivePrefix = {arXiv},
       eprint = {1701.00556},
 primaryClass = {astro-ph.SR},
       adsurl = {https://ui.adsabs.harvard.edu/abs/2017PASP..129d4204M}
}

@ARTICLE{szkody24,
       author = {{Szkody}, Paula and {van Roestel}, Jan and {Mason}, Paul A. and {Littlefield}, Colin and {Rich}, R. Michael and {Bellm}, Eric C. and {Romanov}, Filipp D. and {Healy}, Brian F. and {Jegou du Laz}, Theophile and {Laher}, Russ R. and {Rusholme}, Ben},
        title = "{Spectroscopic Follow-up on Potential Magnetic Cataclysmic Variables}",
      journal = {\aj},
         year = 2024,
        month = may,
       volume = {167},
       number = {5},
          eid = {186},
        pages = {186},
          doi = {10.3847/1538-3881/ad2fcd},
       adsurl = {https://ui.adsabs.harvard.edu/abs/2024AJ....167..186S}
}

@ARTICLE{pagnotta16,
       author = {{Pagnotta}, Ashley and {Zurek}, David},
        title = "{Non-detection of nova shells around asynchronous polars}",
      journal = {\mnras},
         year = 2016,
        month = may,
       volume = {458},
       number = {2},
        pages = {1833-1838},
          doi = {10.1093/mnras/stw424},
archivePrefix = {arXiv},
       eprint = {1603.09370},
 primaryClass = {astro-ph.SR},
       adsurl = {https://ui.adsabs.harvard.edu/abs/2016MNRAS.458.1833P}
}

@ARTICLE{patterson94,
       author = {{Patterson}, Joseph},
        title = "{The DQ Herculis Stars}",
      journal = {\pasp},
         year = 1994,
        month = mar,
       volume = {106},
        pages = {209},
          doi = {10.1086/133375},
       adsurl = {https://ui.adsabs.harvard.edu/abs/1994PASP..106..209P}
}

@ARTICLE{pavlenko_v1500cyg,
       author = {{Pavlenko}, E.~P. and {Mason}, P.~A. and {Sosnovskij}, A.~A. and
         {Shugarov}, S. Yu and {Babina}, Ju V. and {Antonyuk}, K.~A. and
         {Andreev}, M.~V. and {Pit}, N.~V. and {Antonyuk}, O.~I. and
         {Baklanov}, A.~V.},
        title = "{Asynchronous polar V1500 Cyg: orbital, spin, and beat periods}",
      journal = {\mnras},
         year = 2018,
        month = sep,
       volume = {479},
       number = {1},
        pages = {341-347},
          doi = {10.1093/mnras/sty1494},
archivePrefix = {arXiv},
       eprint = {1806.03610},
 primaryClass = {astro-ph.SR},
       adsurl = {https://ui.adsabs.harvard.edu/abs/2018MNRAS.479..341P}
}

@ARTICLE{rawat22,
       author = {{Rawat}, Nikita and {Pandey}, J.~C. and {Joshi}, Arti and {Scaringi}, Simone and {Yadava}, Umesh},
        title = "{SWIFT J0503.7-2819: A nearly synchronous intermediate polar below the period gap?}",
      journal = {\mnras},
         year = 2022,
        month = sep,
          doi = {10.1093/mnras/stac2723},
archivePrefix = {arXiv},
       eprint = {2209.11141},
 primaryClass = {astro-ph.HE},
       adsurl = {https://ui.adsabs.harvard.edu/abs/2022MNRAS.tmp.2511R}
}

@ARTICLE{tess,
       author = {{Ricker}, George R. and {Winn}, Joshua N. and {Vanderspek}, Roland and {Latham}, David W. and {Bakos}, G{\'a}sp{\'a}r {\'A}. and {Bean}, Jacob L. and {Berta-Thompson}, Zachory K. and {Brown}, Timothy M. and {Buchhave}, Lars and {Butler}, Nathaniel R. and {Butler}, R. Paul and {Chaplin}, William J. and {Charbonneau}, David and {Christensen-Dalsgaard}, J{\o}rgen and {Clampin}, Mark and {Deming}, Drake and {Doty}, John and {De Lee}, Nathan and {Dressing}, Courtney and {Dunham}, Edward W. and {Endl}, Michael and {Fressin}, Francois and {Ge}, Jian and {Henning}, Thomas and {Holman}, Matthew J. and {Howard}, Andrew W. and {Ida}, Shigeru and {Jenkins}, Jon M. and {Jernigan}, Garrett and {Johnson}, John Asher and {Kaltenegger}, Lisa and {Kawai}, Nobuyuki and {Kjeldsen}, Hans and {Laughlin}, Gregory and {Levine}, Alan M. and {Lin}, Douglas and {Lissauer}, Jack J. and {MacQueen}, Phillip and {Marcy}, Geoffrey and {McCullough}, Peter R. and {Morton}, Timothy D. and {Narita}, Norio and {Paegert}, Martin and {Palle}, Enric and {Pepe}, Francesco and {Pepper}, Joshua and {Quirrenbach}, Andreas and {Rinehart}, Stephen A. and {Sasselov}, Dimitar and {Sato}, Bun'ei and {Seager}, Sara and {Sozzetti}, Alessandro and {Stassun}, Keivan G. and {Sullivan}, Peter and {Szentgyorgyi}, Andrew and {Torres}, Guillermo and {Udry}, Stephane and {Villasenor}, Joel},
        title = "{Transiting Exoplanet Survey Satellite (TESS)}",
      journal = {Journal of Astronomical Telescopes, Instruments, and Systems},
         year = 2015,
        month = jan,
       volume = {1},
          eid = {014003},
        pages = {014003},
          doi = {10.1117/1.JATIS.1.1.014003},
       adsurl = {https://ui.adsabs.harvard.edu/abs/2015JATIS...1a4003R}
}

@ARTICLE{stockman88,
       author = {{Stockman}, H.~S. and {Schmidt}, Gary D. and {Lamb}, D.~Q.},
        title = "{V1500 Cygni: Discovery of a Magnetic Nova}",
      journal = {\apj},
         year = 1988,
        month = sep,
       volume = {332},
        pages = {282},
          doi = {10.1086/166652},
       adsurl = {https://ui.adsabs.harvard.edu/abs/1988ApJ...332..282S}
}

@ARTICLE{schmidt91,
       author = {{Schmidt}, Gary D. and {Stockman}, H.~S.},
        title = "{Synchronization of the Magnetic Nova V1500 Cygni}",
      journal = {\apj},
         year = 1991,
        month = apr,
       volume = {371},
        pages = {749},
          doi = {10.1086/169939},
       adsurl = {https://ui.adsabs.harvard.edu/abs/1991ApJ...371..749S}
}

@INPROCEEDINGS{lbc_speziali,
       author = {{Speziali}, R. and {Di Paola}, A. and {Giallongo}, E. and {Pedichini}, F. and {Ragazzoni}, R. and {Testa}, V. and {Baruffolo}, A. and {De Santis}, C. and {Diolaiti}, E. and {Farinato}, J. and {Fontana}, A. and {Gallozzi}, S. and {Gasparo}, F. and {Gentile}, G. and {Grazian}, A. and {Manzato}, P. and {Pasian}, F. and {Smareglia}, R. and {Vernet}, E.},
        title = "{The Large Binocular Camera: description and performances of the first binocular imager}",
    booktitle = {Ground-based and Airborne Instrumentation for Astronomy II},
         year = 2008,
       editor = {{McLean}, Ian S. and {Casali}, Mark M.},
       series = {Society of Photo-Optical Instrumentation Engineers (SPIE) Conference Series},
       volume = {7014},
        month = jul,
          eid = {70144T},
        pages = {70144T},
          doi = {10.1117/12.790132},
       adsurl = {https://ui.adsabs.harvard.edu/abs/2008SPIE.7014E..4TS}
}

@ARTICLE{takata22,
       author = {{Takata}, J. and {Wang}, X.~F. and {Kong}, A.~K.~H. and {Mao}, J. and {Hou}, X. and {Hu}, C. -P. and {Lin}, L.~C. -C. and {Li}, K.~L. and {Hui}, C.~Y.},
        title = "{Searching for Cataclysmic Variable Stars in Unidentified X-Ray Sources}",
      journal = {\apj},
         year = 2022,
        month = sep,
       volume = {936},
       number = {2},
          eid = {134},
        pages = {134},
          doi = {10.3847/1538-4357/ac8100},
archivePrefix = {arXiv},
       eprint = {2208.01833},
 primaryClass = {astro-ph.SR},
       adsurl = {https://ui.adsabs.harvard.edu/abs/2022ApJ...936..134T}
}

@ARTICLE{atlas,
       author = {{Tonry}, J.~L. and {Denneau}, L. and {Heinze}, A.~N. and {Stalder}, B. and {Smith}, K.~W. and {Smartt}, S.~J. and {Stubbs}, C.~W. and {Weiland}, H.~J. and {Rest}, A.},
        title = "{ATLAS: A High-cadence All-sky Survey System}",
      journal = {\pasp},
         year = 2018,
        month = jun,
       volume = {130},
       number = {988},
        pages = {064505},
          doi = {10.1088/1538-3873/aabadf},
archivePrefix = {arXiv},
       eprint = {1802.00879},
 primaryClass = {astro-ph.IM},
       adsurl = {https://ui.adsabs.harvard.edu/abs/2018PASP..130f4505T}
}

@ARTICLE{tovmassian,
       author = {{Tovmassian}, G. and {Gonz{\'a}lez-Buitrago}, D. and {Thorstensen}, J. and
         {Kotze}, E. and {Breytenbach}, H. and {Schwope}, A. and
         {Bernardini}, F. and {Zharikov}, S.~V. and {Hernandez}, M.~S. and
         {Buckley}, D.~A.~H. and {de Miguel}, E. and {Hambsch}, F. -J. and
         {Myers}, G. and {Goff}, W. and {Cejudo}, D. and {Starkey}, D. and
         {Campbell}, T. and {Ulowetz}, J. and {Stein}, W. and {Nelson}, P. and
         {Reichart}, D.~E. and {Haislip}, J.~B. and {Ivarsen}, K.~M. and
         {LaCluyze}, A.~P. and {Moore}, J.~P. and {Miroshnichenko}, A.~S.},
        title = "{IGR J19552+0044: A new asynchronous short period polar. Filling the gap between intermediate and ordinary polars}",
      journal = {\aap},
         year = 2017,
        month = dec,
       volume = {608},
          eid = {A36},
        pages = {A36},
          doi = {10.1051/0004-6361/201731323},
archivePrefix = {arXiv},
       eprint = {1710.02126},
 primaryClass = {astro-ph.SR},
       adsurl = {https://ui.adsabs.harvard.edu/abs/2017A&A...608A..36T}
}

@ARTICLE{wang2020,
       author = {{Wang}, Qishan and {Qian}, Shengbang and {Han}, Zhongtao and {Fang}, Xiaohui and {Zang}, Lei and {Liu}, Wei},
        title = "{Spot Model for Identifications of Periods in Asynchronous Polars}",
      journal = {\apj},
         year = 2020,
        month = mar,
       volume = {892},
       number = {1},
          eid = {38},
        pages = {38},
          doi = {10.3847/1538-4357/ab7759},
archivePrefix = {arXiv},
       eprint = {2011.13123},
 primaryClass = {astro-ph.SR},
       adsurl = {https://ui.adsabs.harvard.edu/abs/2020ApJ...892...38W}
}

@BOOK{Warner1995cvs..book.....W,
       author = {{Warner}, Brian},
        title = "{Cataclysmic variable stars}",
         year = 1995,
       volume = {28},
       adsurl = {https://ui.adsabs.harvard.edu/abs/1995cvs..book.....W}
}

@ARTICLE{watson,
       author = {{Watson}, C.~L. and {Henden}, A.~A. and {Price}, A.},
        title = "{The International Variable Star Index (VSX)}",
      journal = {Society for Astronomical Sciences Annual Symposium},
         year = 2006,
        month = may,
       volume = {25},
        pages = {47},
       adsurl = {https://ui.adsabs.harvard.edu/abs/2006SASS...25...47W}
}

@ARTICLE{witham,
       author = {{Witham}, A.~R. and {Knigge}, C. and {Drew}, J.~E. and {Greimel}, R. and {Steeghs}, D. and {G{\"a}nsicke}, B.~T. and {Groot}, P.~J. and {Mampaso}, A.},
        title = "{The IPHAS catalogue of H{\ensuremath{\alpha}} emission-line sources in the northern Galactic plane}",
      journal = {\mnras},
         year = 2008,
        month = mar,
       volume = {384},
       number = {4},
        pages = {1277-1288},
          doi = {10.1111/j.1365-2966.2007.12774.x},
archivePrefix = {arXiv},
       eprint = {0712.0988},
 primaryClass = {astro-ph},
       adsurl = {https://ui.adsabs.harvard.edu/abs/2008MNRAS.384.1277W}
}

@ARTICLE{wynn92,
       author = {{Wynn}, G.~A. and {King}, A.~R.},
        title = "{Theoretical X-ray power spectra of intermediate polars}",
      journal = {\mnras},
         year = 1992,
        month = mar,
       volume = {255},
        pages = {83-91},
          doi = {10.1093/mnras/255.1.83},
       adsurl = {https://ui.adsabs.harvard.edu/abs/1992MNRAS.255...83W}
}

@ARTICLE{zucker95,
       author = {{Zucker}, D.~B. and {Raymond}, J.~C. and {Silber}, A. and {Mason}, P. and {Curiel}, S. and {Vrtilek}, S. and {Schlegel}, E.},
        title = "{Phase-resolved IUE and Optical Observations of the Polar BY Camelopardalis}",
      journal = {\apj},
         year = 1995,
        month = aug,
       volume = {449},
        pages = {310},
          doi = {10.1086/176056},
       adsurl = {https://ui.adsabs.harvard.edu/abs/1995ApJ...449..310Z}
}

\appendix

In the VSX catalog, two objects classified as polars have orbital frequencies above that of \gaia: ZTF19aasmeay and Gaia~19bxc \citep{gaia19bxc, Galiullin}, with orbital frequencies of 18.85~\cpd\ and 22.35~\cpd, respectively. Intrigued, we examined the TESS photometry for these objects.

Our analysis of the TESS FFI observations of ZTF19aasmeay from sector 81 shows a single frequency at 17.85~\cpd, which we presume to be the orbital frequency. It is exactly 1~\cpd\ lower than the nominal frequency in the VSX catalog, which used ground-based photometry to measure the orbital frequency. We conclude that sampling-related aliases led to a small misidentification in the orbital frequency and that the 17.85~\cpd\ signal in TESS is the true orbital frequency.

On the other hand, the TESS FFI data confirm Gaia19bxc's remarkably short period, first detected by \citet{gaia19bxc} and confirmed photometrically and spectroscopically by the detailed study in \citet{Galiullin}. The TESS data show both the fundamental frequency and its second harmonic and are fully consistent with \citet{Galiullin}. \footnote{In the original version of the preprint of this manuscript, we speculated that Gaia19bxc might be an asynchronous mCV similar to \ztf, with a spin period of 66~min and a longer, undetected orbital period above the period minimum. However, \citet{Galiullin} found no evidence of asynchronous rotation, so our original musings about this scenario appear to be untenable. Instead, the available data suggest that Gaia19bxc is a synchronous polar below the period minimum. We thank Vladislav Dodon for alerting us to the \citet{Galiullin} paper.}

\begin{figure}
    \includegraphics[width=\columnwidth]{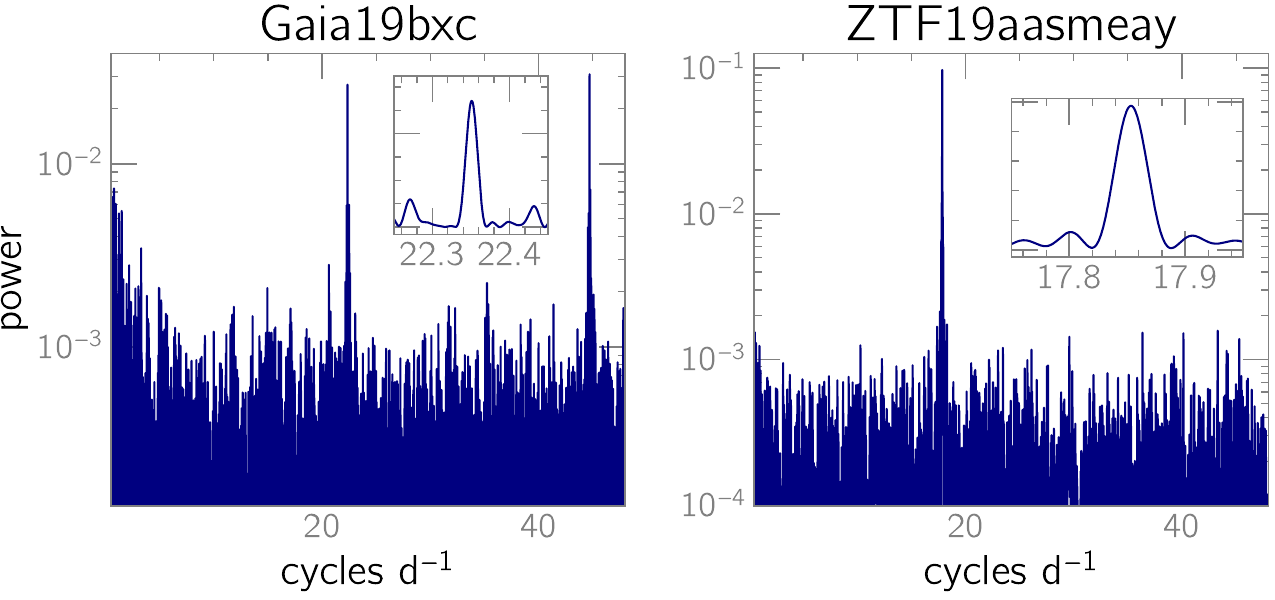}
    \caption{TESS power spectra of Gaia19bxc and ZTF19aasmeay, two understudied systems classified as very-short-period polars in the VSX catalog. The TESS data for Gaia19bxc show the likely orbital frequency from \citet{Galiullin} and its next harmonic. Conversely, the TESS data for ZTF19aasmeay do not show the putative orbital frequency from the VSX (18.85~\cpd) and instead show a single frequency at 17.85~\cpd. The 1~\cpd\ difference suggests that ground-based sampling aliases led to the selection of the incorrect daily alias of the true signal. }
    \label{fig:enter-label}
\end{figure}

\end{document}